\documentclass[usenatbib]{mn2e}

\usepackage{arydshln}
\usepackage{amssymb}
\usepackage{relsize}
\usepackage{ragged2e}
\usepackage{subfig}
\usepackage{bm}

\usepackage{amsmath}
\usepackage{graphicx}
\usepackage{color}
\usepackage{aas_macros}
\usepackage{float}

\newcommand{\width}{0.5\textwidth}
\newcommand{\tserieslong}{Combining spectroscopic and photometric surveys
using angular cross-correlations}
\newcommand{\tseriesshort}{Combining spectroscopic and photometric surveys}

\newcommand{\tbias}{Galaxy bias and stochastisity}

\newcommand{\pssky}{EG14a}

\newcommand{\gkone}{FxB-$\left<\delta_F \gamma_F\right>$}
\newcommand{\gktwo}{FxB-$\left<\delta_B \gamma_F\right>$}
\newcommand{\gkthree}{FxB-$\left<\delta \: \gamma \right>$}

\newcommand{\gkfb}{FxB-$\left<\text{FB}\right>$}

\newcommand{\nbright}{71}

\newcommand{\be}{\begin{equation}}
\newcommand{\ee}{\end{equation}}
\newcommand{\xbf}{\begin{figure}}
\newcommand{\xef}{\end{figure}}
\newcommand{\xbfd}{\begin{figure*}}
\newcommand{\xefd}{\end{figure*}}
\newcommand{\xbt}{\begin{table}}
\newcommand{\xet}{\end{table}}

\newcommand{\citedir}{\citet}
\newcommand{\citeind}{\citep}

\newcommand{\tmpcitetwo}[1]{}
{}
\newcommand{\tmprm}[1]{}
\newcommand{\xfigure}[1]{
\begin{center}
\includegraphics[width=\width]{links/#1}
\end{center}
}

\newcommand{\zfigure}[1]{
\begin{center}
\includegraphics[width=1.0\textwidth]{links/#1}
\end{center}
}

\newcommand{\dzmax}{\Delta Z_ \text{Max}}
\newcommand{\dzbin}{\Delta z}

\newcommand{\sn}{S/N }
\newcommand{\sigmaz}{\sigma_z}
\newcommand{\sqarcmin}{\text{arcmin}^2}

\newcommand{\fom}{\text{FoM}_{w}}
\newcommand{\fomg}{\text{FoM}_\gamma}
\newcommand{\fomc}{\text{FoM}_{w\gamma}}
\newcommand{\fomdetf}{\text{FoM}_{\text{DETF}}}

\newcommand{\citeoverlap}{\citeind{cai2011,gazta,cai2012,kirk,mcdonald,deputter}}

\begin{document}

\title[\tseriesshort\ III: \tbias]{\tserieslong\ III: \tbias}
\author[Martin Eriksen, Enrique Gazta\~naga]
{\parbox{\textwidth}{Martin Eriksen$^{1,2}$, Enrique Gazta\~naga$^1$}
\vspace{0.4cm}\\
$^1$Institut de Ci\`encies de l'Espai (IEEC-CSIC),  E-08193 Bellaterra (Barcelona), Spain \\
$^2$Leiden Observatory, Leiden University, PO Box 9513, NL-2300 RA Leiden, Netherlands \\
}

\maketitle
\begin{abstract}
In the first paper of this series, we studied the effect of baryon acoustic
oscillations (BAO), redshift space distortions (RSD) and weak lensing (WL) on
measurements of angular cross-correlations in narrow redshift bins. Paper-II
presented a multitracer forecast as Figures of Merit (FoM), combining a
photometric and spectroscopic stage-IV survey. The uncertainties from galaxy
bias, the way light traces mass, is an important ingredient in the forecast.
Fixing the bias would increase our FoM equivalent to 3.3 times larger area for
the combined constraints. This paper focus on how the modelling of bias affect
these results. In the combined forecast, lensing both help and benefit from
the improved bias measurements in overlapping surveys after marginalizing
over the cosmological parameters. Adding a second lens population in
counts-shear does not have a large impact on bias error, but removing all
counts-shear information increases the bias error in a significant way. We also
discuss the relative impact of WL, magnification, RSD and BAO, and how results
change as a function of bias amplitude, photo-z error and sample density. By
default we use one bias parameter per bin (with 72 narrow bins), but we show
that the results do not change much when we use other parameterizations, with
at least 3 parameters in total. Bias stochasticity, even when added as one new
free parameter per bin, only produce moderate decrease in the FoM. In general,
we find that the degradation in the figure of merit caused by the uncertainties
in the knowledge of bias is significantly smaller for overlapping surveys.

\end{abstract}

\section{Introduction}
The late time expansion of the Universe can be measured through various
probes, including super novas \citeind{riessSN,perlmutterSN}, the cosmic
microwave background (CMB) \citeind{planck2015}, weak lensing shear
\citeind{heymans_lens} and galaxy clustering \citeind{andersonboss}.
Galaxy clustering is assumed and measured to directly trace the underlying dark
matter distribution \citeind{clustering_kaiser, clustering_bardeen}, but
require measuring the galaxy bias. The weak gravitational lensing directly
probe the mass. However, the lensing kernel is broad, which reduce its ability
to measure the growth.

Weak lensing (WL) survey relies on averaging a large sample of galaxies, which
is only possible in photometric surveys with photo-z techniques. Ongoing
surveys include 
DES\citeind{des}\footnote{http://www.darkenergysurvey.org/} and
KiDS\citeind{kids}\footnote{http://kids.strw.leidenuniv.nl/}, while 
Euclid\citeind{euclid}\footnote{http://www.euclid-ec.org/} and 
LSST\citeind{lsst1}\footnote{http://www.lsst.org/}
are next generation WL surveys. The galaxy clustering relies on
accurate redshift to measure the BAO peak position and
the radial clustering. Past and ongoing redshift surveys include VVDS
\citeind{vvds}\footnote{http://cesam.oamp.fr/vvdsproject/}, VIPERS
\citeind{vipers}\footnote{http://vipers.inaf.it/} and WiggleZ
\citeind{wigglez}\footnote{http://wigglez.swin.edu.au/site/}, while DESI
\citeind{msdesi}\footnote{http://desi.lbl.gov/} is expected to start in 2018.

The galaxy clustering and weak lensing are complimentary, probing different
parameter degeneracies \citeind{simpsonWLRSD}, being subject to different
experimental difficulties and astrophysical biases \citeind{weinberg}. How
should photometric and spectroscopic optimally be combined? Several groups have
investigate the benefit of overlapping surveys with varying conclusions (see
paper-II) \citeoverlap. Our three article series investigates the combined
constraints, focusing on the same-sky benefit. In \citedir{paperI} (from now
on paper-I) we studied the algorithm and modelling of cross-correlations in
narrow redshift bins. \citedir{paperII} (from now on paper-II) presented the
forecast, while \citedir{sameskyX} (from now on \pssky) companion paper
elaborated on the same-sky benefit. This paper (paper-III) investigate the 
role of the galaxy bias and stochasticity.

While the galaxy and dark matter distribution is related, the exact relation
depends on galaxy formation \citeind{galform}, galaxy evolution
\citeind{galaxyev1, galaxyev2, galaxyev3, galaxyev_blanton} and selection
effects. For constraining cosmology with galaxy distributions, we need to
model galaxy bias and marginalize over uncertainties in the modelling. To
reduce requirements on modelling bias, we can constrain
cosmology from the excess of galaxy pairs with 150 Mpc separation, the BAO peak
\citeind{eisenstein_bao1, eisenstein_bao2, anderson_bao}. In these papers,
constraints from galaxy clustering and galaxy-shear cross-correlations use the
full correlation function, which require modelling the galaxy bias
\citeind{fullcls}. This improve the parameter constraints, but the bias becomes
an important part of the forecast and is therefore studied here in detail.

Overlapping galaxy surveys allows directly cross-correlation the galaxy
samples. Additionally, the sample variance reduce through both galaxy
populations tracing the same matter fluctuations \citeind{mcdonseljak}.
Paper-II showed that both effects contribute about equal to the
combined constraints and \pssky\ stressed the importance of the
covariance between the spectroscopic and photometric sample. This paper
also investigate the additional correlations and the covariance from
overlapping volumes, but present the forecasted bias error. In particular, we
focus on the counts-shear cross-correlations, since the contribution was
largely independent on including lenses from one or both surveys.

The HOD model \citeind{hod1,zhenghod,couponhod,cacciato_hod} predicts the galaxy
bias from physical assumptions on how galaxies occupy dense regions, which can
be implemented in simulations and tested against real data. The fiducial bias
is motivated by a simple HOD model that shows a scale-independent and linear
bias at large scales \citeind{gazta}. While the bias evolution in a linear
model can be parameterized with a few parameters, the fiducial parameterization
use one parameter for each redshift bin. Section \ref{secbias:prandmodel}
therefore compare the one-bias-per-bin forecast with a model using a linear
interpolation between pivot points.

The fiducial bias model, $\delta_g = b(z,k) \delta_m$, relate linearly and
deterministic the galaxy ($\delta_g$) and matter ($\delta_m$)
overdensities. In reality, the relation at small scales include at stochastic
component \citeind{biasstoch}. The stochastic component change the
auto-correlation of galaxy counts and cancelling in the counts-shear
cross-correlations. While the linear scale included in the forecast avoid the
bias stochasticity, we study the impact using a simple model (see subsections
\ref{thr_stoch}, \ref{fombias_stoch}). In addition to reducing the
signal-to-noise, one need to marginalize over the bias stochasticity model
parameters. The overlapping surveys is of particular interest, since the
overlap can provide additional constraints to these parameters.

This paper is organized in the following manner. Section \ref{secbias:error} 
presents the forecast assumptions, bias derivative formula, fiducial errors
and how counts-shear and the sample variance cancellation contribute to
the bias error. Then section \ref{secbias:prandmodel} looks at bias priors, 
bias parameterization and tests the impact of a stochastic bias. Appendix
\ref{app:amplitude} study how shifting the bias amplitude change the forecast
and the relative RSD, BAO and WL contributions. Finally appendix
\ref{app:errsurvey} study how the galaxy density and redshift uncertainties
change the bias errors.

\section{The bias constraints}
\label{secbias:error}
\subsection{Forecast assumptions and notation}
\label{subsec:assumpt}
\newcommand{\tblskipa}{\noalign{\vskip 1.0mm}}
\newcommand{\tblmidlinea}{\tblskipa\hdashline\tblskipa}

\begin{table}
\begin{center}
\vskip+0.5ex
\begin{tabular}{lrr}
\hline
Parameter & Photometric (F) & Spectroscopic (B)\\
\hline
%\midrule
Area [sq.deg.] & 14,000 & 14,000 \\
Magnitude limit & $i_{AB} < 24.1$& $i_{AB} < 22.5$ \\
Redshift range & $0.1 < z < 1.5$ & $0.1 < z < 1.25$ \\
Redshift uncertainty & 0.05(1+z) & 0.001(1+z) \\
z Bin width & 0.07(1+z)& 0.01(1+z) \\
Number of bins & 12 & \nbright \\
Density [gal/arcmin$^2$] &  6.5 & 0.4 \\
\hline
%\bottomrule
\end{tabular}
\end{center}
\caption{Parameters describing the assumed photometric and spectroscopic
stage-IV surveys in the forecast. See paper-II and \pssky\ 
for further details.}
\label{tbl_assumpt}
\end{table}

Paper-I described the formalism for galaxy clustering, RSD and weak lensing in
2D correlations. In paper-II we presented the Fisher formalism, notation,
forecast assumptions and Figures of Merit (FoM). This subsection include, to
improve readability, the central forecast setup and notation. The reader
is referred to paper-II for a full description.

The forecast define a photometric(F/Faint) and a spectroscopic(B/Bright) 
stage-IV survey \citeind{detf}. Table \ref{tbl_assumpt} summarize the most
important parameters and the forecast use Fisher matrices to estimate the bias
errors and FoMs. The fiducial bias redshift evolution is given by

\begin{align}
b_F(z) &= 1.2 + 0.4 (z - 0.5) \\
b_B(z) &= 2.0 + 2.0 (z - 0.5)
\label{fidbias}
\end{align}

\noindent
while the default bias parameterization use one parameter per redshift bin and
population. The observables included are auto- and cross-correlations between
galaxy counts ($\delta$) and shear ($\gamma$). The redshift space distortions
(RSD) and radial information in the galaxy sample is measured by using narrow
redshift bins for the spectroscopic sample \citeind{rsdin2d,asorey,dionbins,asorey2}.

These papers focus on the benefit of combining spectroscopic and photometric
surveys and the notation FxB is short for F and B being overlapping
surveys on the sky, while F+B means non-overlapping surveys. The default
forecast include both galaxy counts and shear (All), but some result only
include galaxy counts (Counts). The Figure of Merit (FoM) is defined by

\be
\text{FoM} = \frac{1}{\sqrt{\det(S)}}
\ee

\noindent
where $S$ is a subspace of parameter the Fisher forecasted covariance
matrix. For $\fomdetf$, the Dark Energy Task Force (DETF) FoM, then 
$S=[w_0, wa]$. The $\fomg$ is the inverse error of $\gamma$, so that the covariance subspace
(S) only include $\gamma$. For $\fomc$ then S includes $w_0$, $w_a$, $\gamma$
and measure if a probe can constrain the combined expansion and growth history.
Unless explicitly stated, the FoMs and bias errors marginalize over $w_0, w_a,
h, n_s, \Omega_m, \Omega_b, \Omega_{DE}, \sigma_8, \gamma$. Planck
priors\footnote{http://www.physics.ucdavis.edu/DETFast/}is always included.

\subsection{Derivative of the galaxy bias}
\label{subsec_biasder}
The default bias parameterization is one parameter per redshift bin and galaxy
population. This subsection presents a formula for the derivative of the 2D
correlation with the bias when including photo-z, RSD, magnification and
multiple galaxy populations.

Photo-z uncertainties cause overdensities with origin at one redshift, to be
observed at another redshift. One can express the observed fluctuations as a
linear contribution from different redshifts. The galaxy bias 
evolve with redshift and the observed overdensity is a convolution of the bias
and redshift selection function (see paper-I). For the bias derivative, the
parameters can either be defined as the true bias or an effective bias after
including the photo-z effect. These papers (paper-II, paper-III, \pssky), use
the true bias as nuisance parameters.

\newcommand{\mtx}[1]{\bm{\mathsf{#1}}}
The photo-z effect in 2D correlations can be approximated by the transition
matrices \citeind{gazta}. Let $\mtx{C}$ be a matrix of 2D correlation predictions in
top-hat bins. The observed correlations $\mtx{\tilde{C}}$ including photo-z effects
can be written

\be
\tilde{C}_{ij} = \sum_{mn} r_{im} r_{jn} C_{mn}
\label{trans_mat}
\ee

\noindent
where $r$ is a transition matrix. The matrix element $r_{ij}$ is the fraction
of a fluctuation in bin $i$ which originates from bin $j$. Summing over the
transition matrix elements is effectively a low resolution integral over the
redshift selection function. In Eq.\ref{trans_mat} the first and second transition
matrix respectively corresponds to the first and second index of \mtx{C}. When
generalizing the formula to several populations and overdensity types (counts,
shear), the two transition matrices (r) will differ. While the transition
matrices for the Bright and Faint populations can be different, they are in
these papers considered equal for shear and galaxy counts for each population
(Faint, Bright).

The transition matrix leads naturally to an expression for the bias derivative.
Let $\tilde{C}_{ij}^{AB}$ be the observed correlation of \mtx{A} and \mtx{B},
in respectively redshift bin $i$ and $j$. Define further $b_y^X$ to be the bias
of overdensity type X in redshift bin y. The derivative of the observed
correlation with respect to the bias is then

\newcommand{\mat}{\rho}
\begin{align}
\frac{\partial \tilde{C}_{ij}^{AB}}{\partial b^X_y} \
=  \sum_{m,n} \left[ r_{iy}^{X} r_{jn}^B C_{yn}^{\mat B} + \
   r_{im}^{A} r_{jy}^X C_{my}^{A \mat} \right]
\label{derbias}
\end{align}

\noindent
where $C^{\mat X}$ is the cross-correlation of dark matter in real space and
one overdensity of type $X$. These papers only include the parameter $b^X$ for
galaxy counts, but Eq.\ref{derbias} can in general be used for shear intrinsic
alignments, clustering of galaxy sizes or clustering of magnitudes.

Galaxy clustering, redshift space distortions, cosmic magnification and other
minor effects, all contribute to the galaxy counts overdensities. Among these
effects, only the galaxy clustering in real space depends on the bias. Most
importantly, the RSD contribution depends on the velocity introduced by the
underlying matter fluctuations and is independent of the galaxy bias. A
central idea in these papers is to measure RSD in 2D correlation functions by
using narrow redshift bins. Since the derivative should only consider parts of
the correlations caused by the galaxy clustering, Eq.\ref{derbias} includes the
cross-correlation between matter and the observable. When using the algorithm
outlined in paper-I, then estimating these cross-correlations does not require
redoing the full calculations, but can be performed faster by reusing
intermediate results in the calculations.

\subsection{Bias errors in the fiducial configuration.}
\label{errbias_fiducial}
\xbf
\xfigure{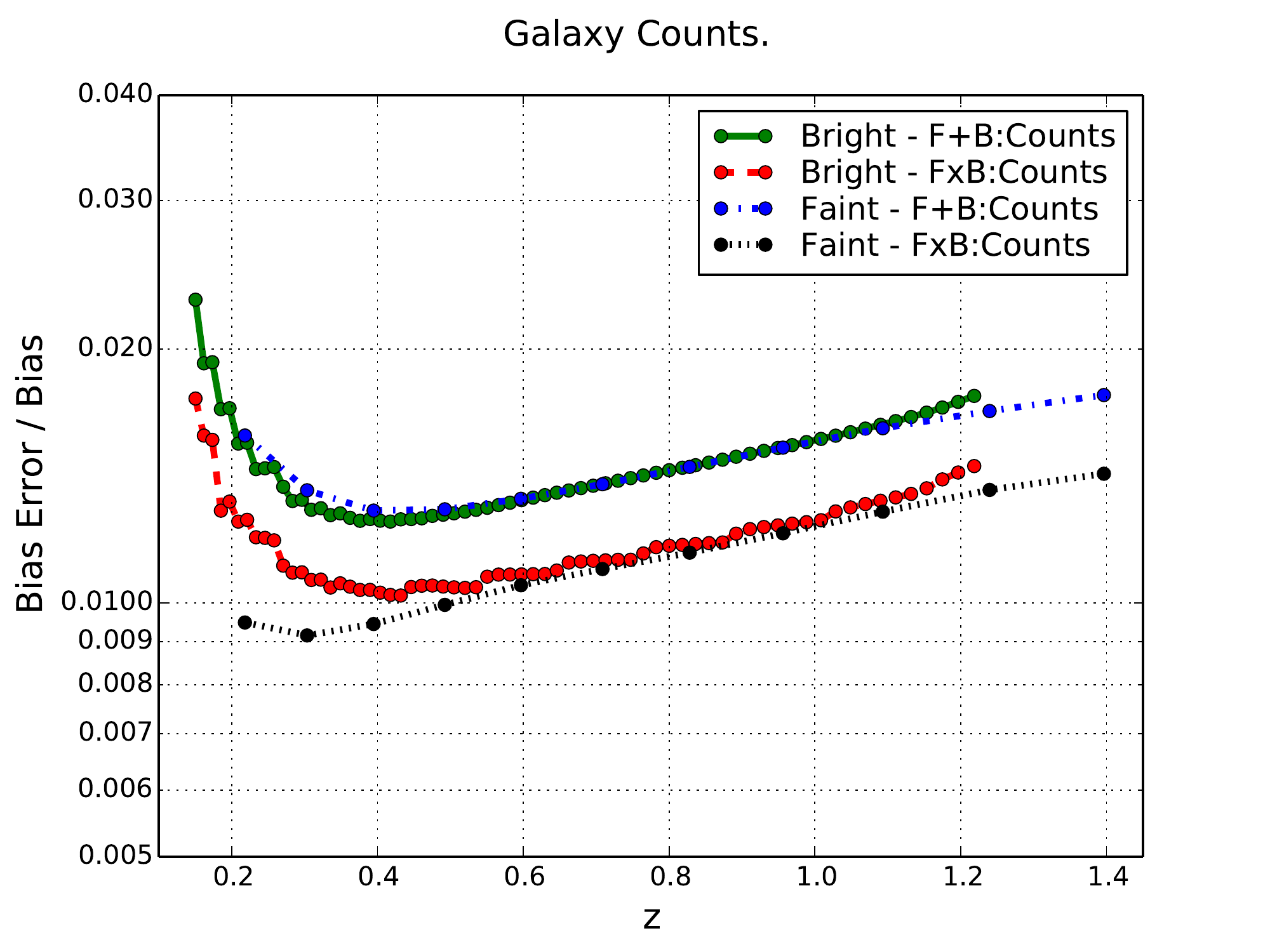}
\xfigure{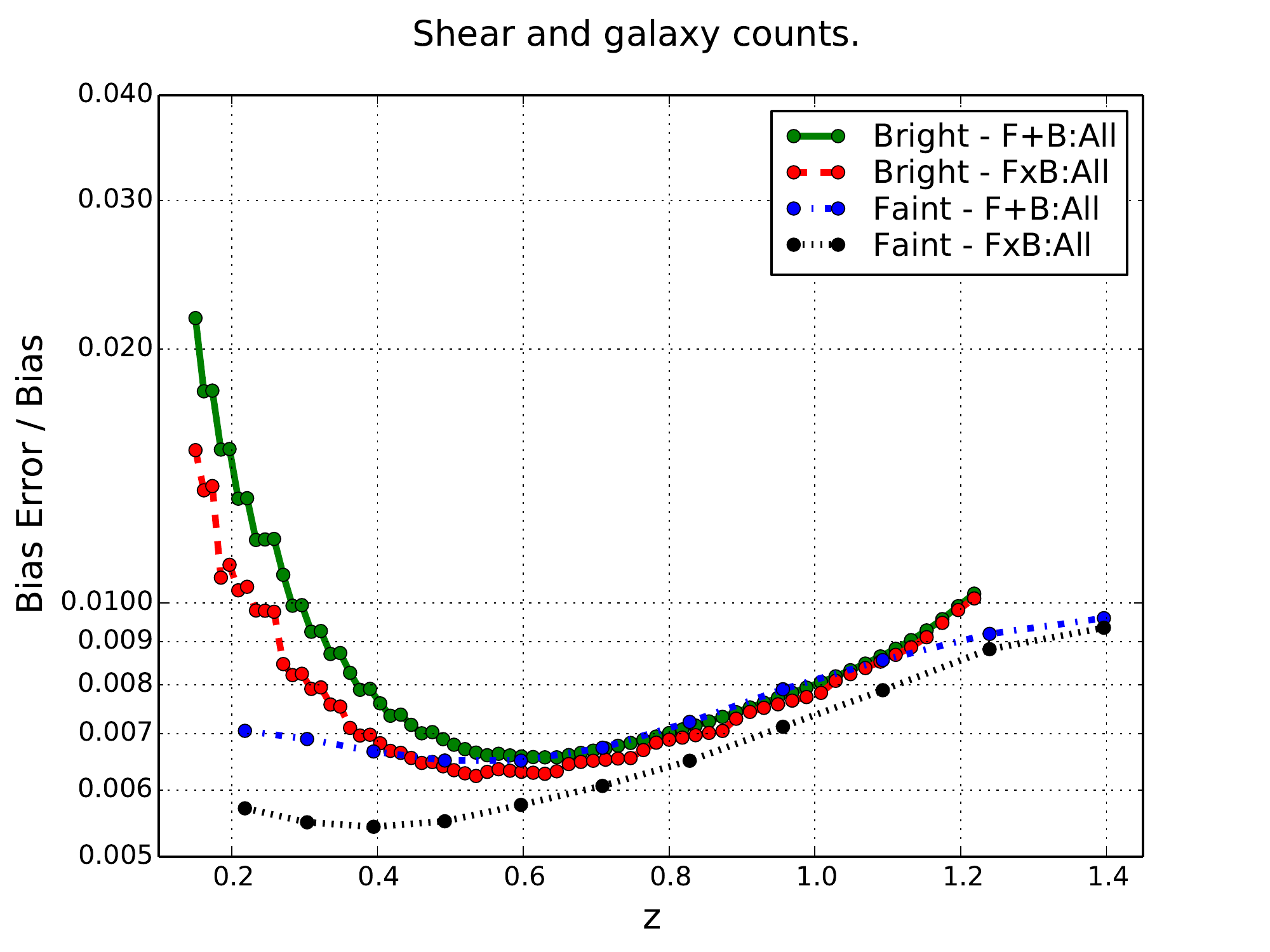}
\caption{The relative error on galaxy bias, marginalized over other cosmological parameters
and including Planck priors. The four lines corresponds to the
Bright and Faint populations, for overlapping (x) and non-overlapping (+)
surveys. Points corresponds to errors measured using one bias parameter
for each bin. On the x-axis is the mean for each the corresponding redshift
bin. For the bright sample the bins are thinner, which results in more points.
In the top panel the measurements only includes galaxy counts, while the lower
panel also includes shear.}
\label{errbias_amp}
\xef

Fig.\ref{errbias_amp} shows the estimated bias errors relative to the fiducial
bias. While the forecast include redshift bins from $z = 0.1$ for both galaxy
populations, the first bins has little constraining power. At low redshifts
the bias error diverges and we therefore show only the error for $z < 0.2$. For
both the exact calculations and the Limber approximation, the main contribution
to an auto-correlation signal is the scale $k = (l+0.1) / \chi(z)$ and lower
redshifts enter into non-linear scales for smaller l. We remove in the forecast
correlations which enters into the non-linear regime (see paper-II) and this
leads to worse constrains at low redshifts. Also, the bias is parameterized
using the underlying/true bias (see subsection \ref{subsec_biasder}). Even if
a redshift bin is excluded at a particular scale (l-value), the bias parameter
enters in cross-correlations measured in nearby bins from the photo-z
dispersion.

Lines in Fig.\ref{errbias_amp} show the relative bias error. The top panel only
include galaxy counts in the forecast. For both populations the estimated bias
error are similar for the Bright and Faint population, except for overlapping
surveys at low redshifts where magnification contribute stronger to the Faint
bias constraints. For galaxy clustering the relative bias constraints are very
similar for both populations, but becomes different if fixing the cosmological
parameters (not shown). Overlapping galaxy surveys significantly lower the bias
and benefit the two populations about equally much. In the lower panel also
counts-shear and shear-shear correlations are included in the forecast. Lensing
both directly measuring bias through the counts-shear observables, but also
indirectly through better constraining cosmological parameters, which are
discussed respectively in subsection \ref{errbias_cross} and
\ref{errbias_cosmo}.

\xbf
\xfigure{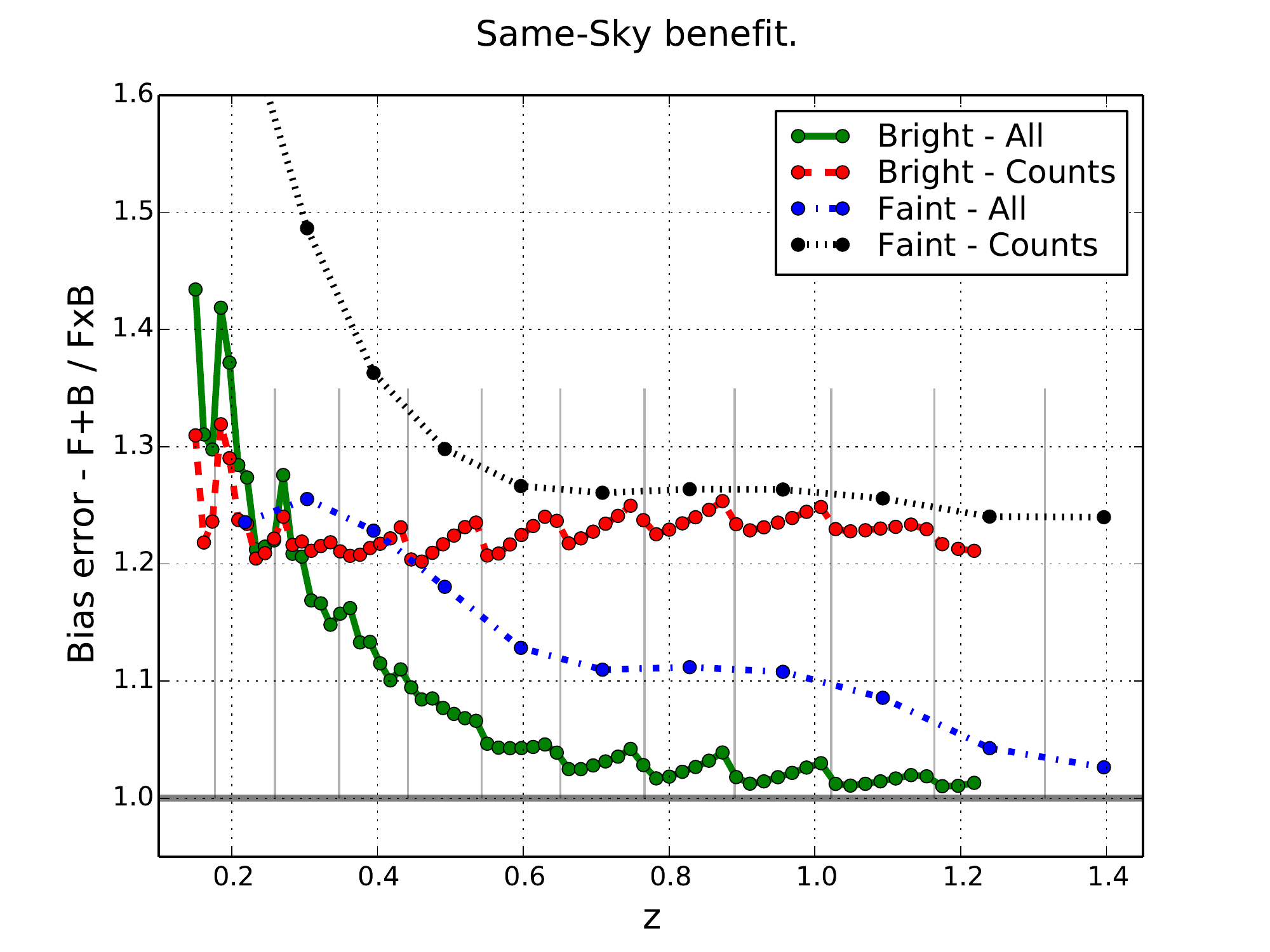}
\caption{The same-sky benefit on measuring the galaxy bias. Lines show the
non-overlapping/overlapping samples bias error ratio. Each line plots the bias
error in non-overlapping samples over overlapping samples. For ratios over
unity overlapping samples are beneficial in measuring the bias. The thin
vertical lines mark the edges of the Faint redshift bins.}
\label{errbias_err_samesky}
\xef

Fig. \ref{errbias_err_samesky} shows directly the F+B/FxB bias error ratios, 
which is the same-sky benefit. For both the Bright and Faint bias, with and
without shear, overlapping surveys leads to better bias constraints. Including
the shear signal reduce the benefit of overlapping samples compared to only
galaxy counts. This trend was previously seen in the forecast (paper-II, Table
4 and 5) and is here shown directly for the bias error. When including lensing,
both the galaxy clustering and counts-shear observables can measure the bias.
While this improve the bias measurements and forecast, the multiple ways of
constraining the bias lower the effect of overlapping Bright and Faint galaxy
counts. The Bright bias error also depend on which Faint redshift bin    the
Bright bin overlap. This cause a small drop in the Bright ratios at the edges
of a Faint bin, which are marked with vertical lines.

\subsection{Correlation between cosmology and bias}
\label{errbias_cosmo}
\xbf
\xfigure{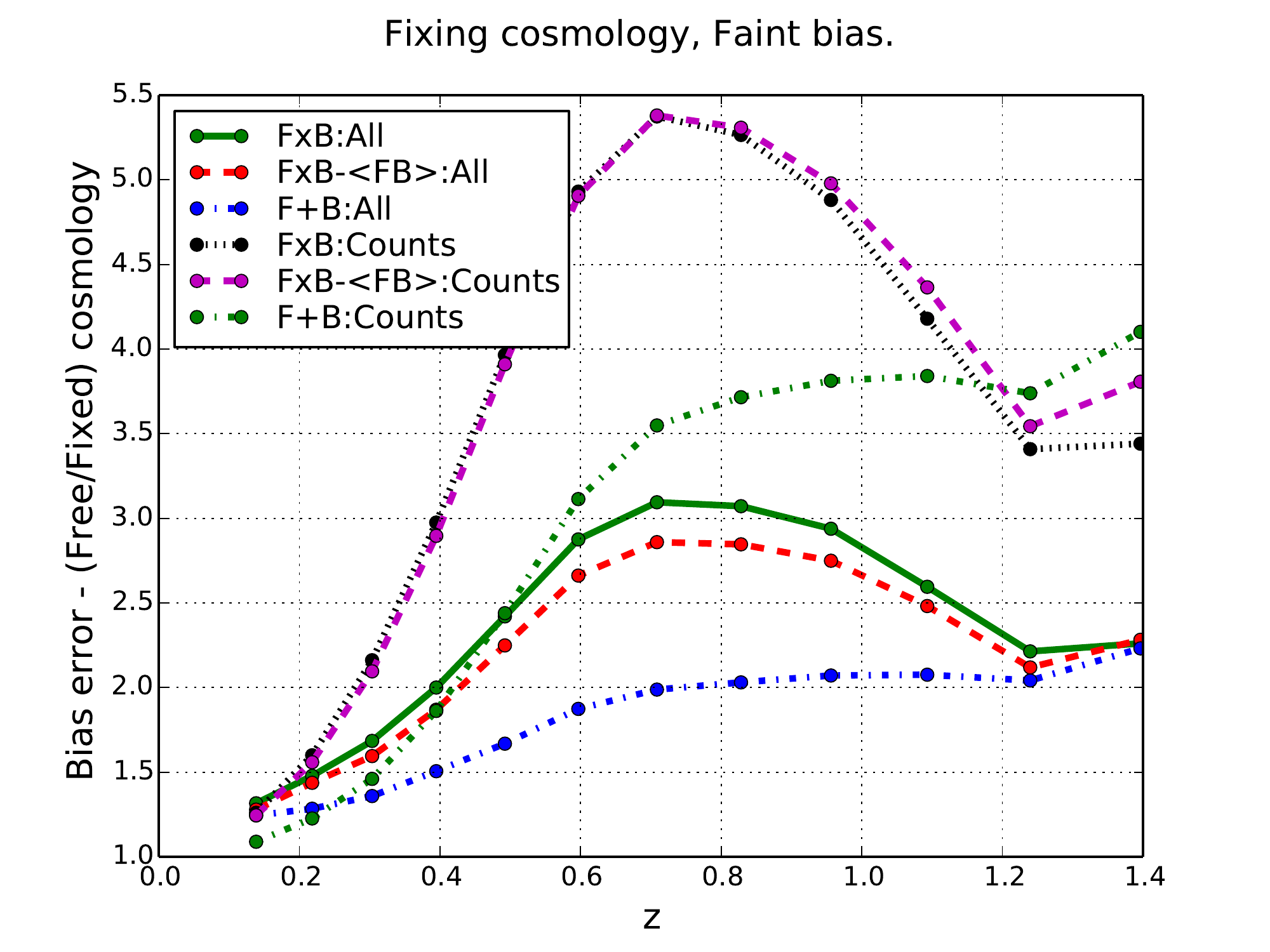}
\caption{Effect of the covariance between cosmological and bias parameters.
The lines show the Faint bias error ratios for free/fixed cosmological parameters. On
the x-axis is the redshift. Two lines (FxB:All, F+B:All) plots the ratio
including galaxy counts and shear, while the other two (FxB:Counts, F+B:Counts)
only include galaxy counts.}
\label{errbias_cosmo_amp}
\xef

The cosmological constraints from galaxy clustering require marginalizing over
uncertainty in the galaxy bias modelling. Fixing the galaxy bias would increase
the $\fomc$ forecast for FxB:All equivalent to a 3.3 larger survey area. The
cosmology constraints depend both on measuring the bias and the lowest possible
covariance between the cosmological and bias/nuisance parameters (see \cite{sameskyX}).
In studying the bias error, one should account for this covariance. This
subsection focus on how marginalizing/fixing cosmological parameters would
affect conclusions on overlapping surveys and the importance of weak lensing.

Fig. \ref{errbias_cosmo_amp} shows the Faint free/fixed cosmology bias error
ratio. These ratios should always be above unity, since marginalizing over
cosmology increase the bias errors. To limit the discussion, the figure and
discussion are for brevity restricted to only the Faint bias. While the
magnitude and exact details differs, the main conclusions are similar for the 
Bright sample. In this figure, the FxB ratios peaks around $z=0.7$, which
results from a peak in the Faint galaxy density and disappear a very high
galaxy density (not shown). Around $z=0.7$, the FxB:Counts(F+B:Counts) ratio
is 5.3(3.1) and this change in bias error corresponds to effectively 28(9)
times larger area. In comparison, FxB:All versus F+B:All only increase $\fomc$
for FxB:All equivalent to 50\% larger area. When studying how improved bias
measurements impact overlapping surveys, one need to account for the covariance
between cosmological and bias parameters. Results without covariance will tend
to underestimate the benefit of overlapping surveys.

\xbf
%\xfigure{prod_biasgr33.pdf}
\xfigure{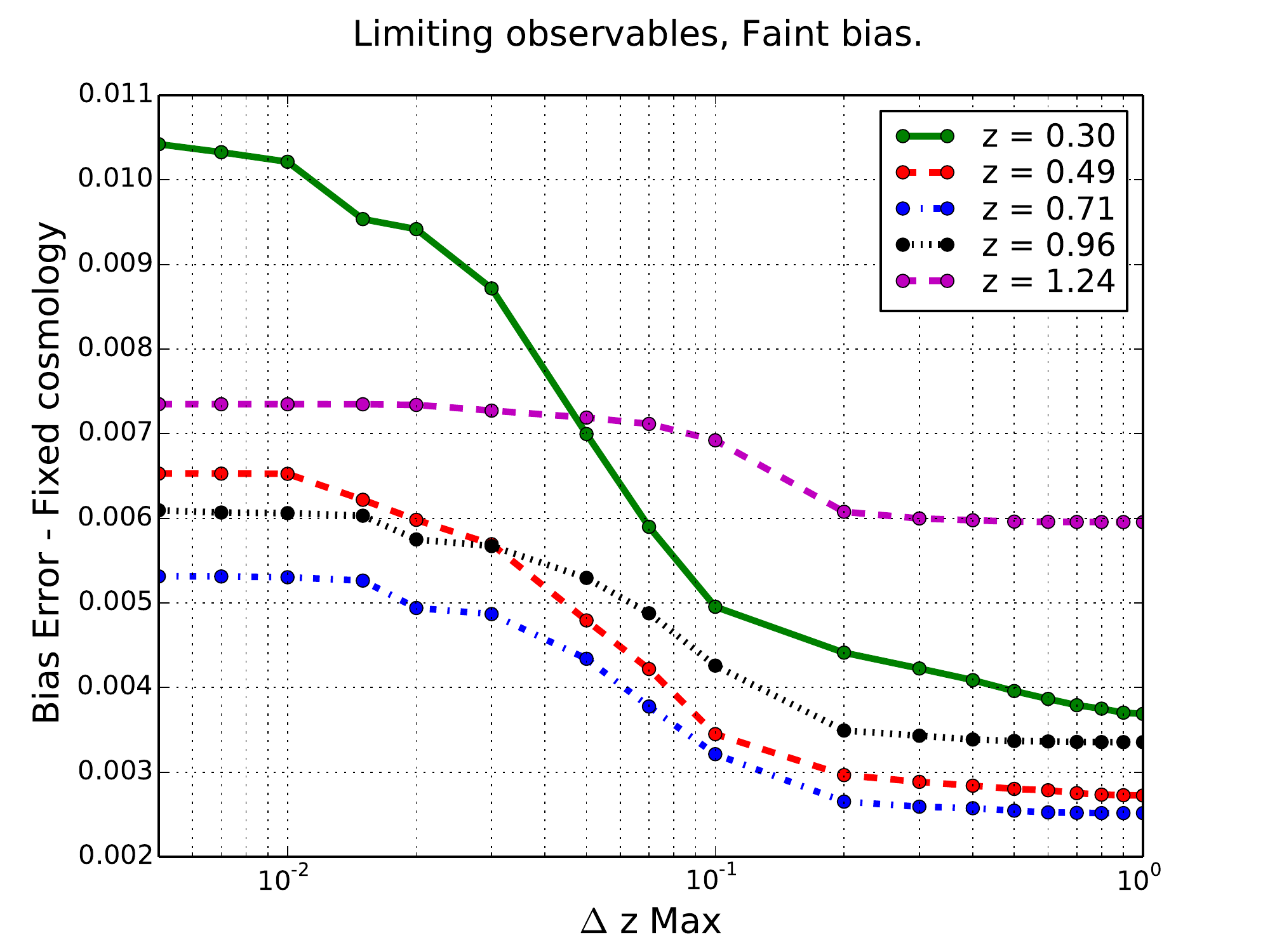}
\xfigure{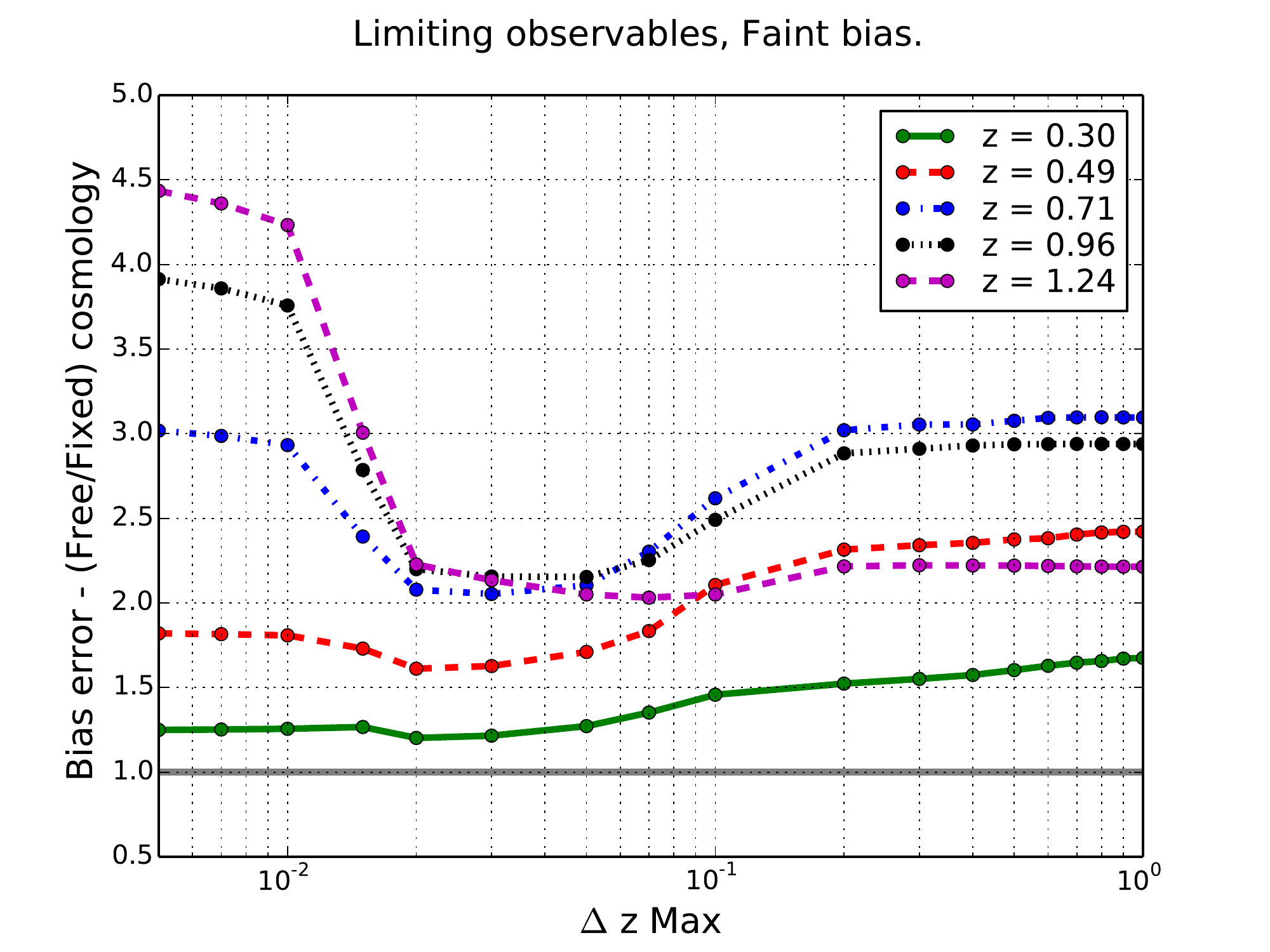}
\caption{
Effect of limiting the redshift separation in the cross-correlations. Errors
are estimated for FxB:All. The top panel shows the bias errors, while the bottom
panel shows the free/fixed cosmology bias error ratio.  On the x-axis is the
maximum distance between the mean of the two redshift bins in a correlation
($\dzmax$).}
\label{errbias_cross_upper}
\xef

Fig.\ref{errbias_cross_upper} shows results when varying $\dzmax$. The $\dzmax$
requirement limit which correlations are included in the forecast. It was
introduced in paper-II and limit the forecast to only include
cross-correlations with ($|z_j - z_i| \leq \dzmax$), where $z_i$ and $z_j$ are
the mean redshifts for the two overdensities in the cross-correlation. The
auto-correlations are always included. For $0.0 < \dzmax < 0.1$ the forecast
also include Bright-Faint galaxy counts cross-correlations, while $0.01 <
\dzmax \lesssim 0.03$ include the important radial (between redshift)
Bright-Bright cross-correlations. The counts-shear correlations are first
important at larger $\dzmax$. For further examples of usage, see paper-II.

The top panel (Fig. \ref{errbias_cross_upper}) show the estimated Faint bias
error, fixing the cosmological parameters. When increasing $\dzmax$, the
Bright-Faint cross-correlations are included and the bias error decline
steadily. At higher $\dzmax$ the counts-shear only leads to very small
improvements in the bias error. In the bottom panels are bias error ratios for
free/fixed cosmology. For $0.01 < \dzmax < 0.02$ the forecast includes
Bright-Bright radial cross-correlations, which measure cosmological parameters
and the ratio drops. At high $\dzmax$ the observables include the counts-shear
cross-correlation. This signal increase the correlation between cosmology and
bias parameters, which results in the ratios increasing.

\subsection{Counts-shear cross-correlations}
\label{errbias_cross}

\xbf
\xfigure{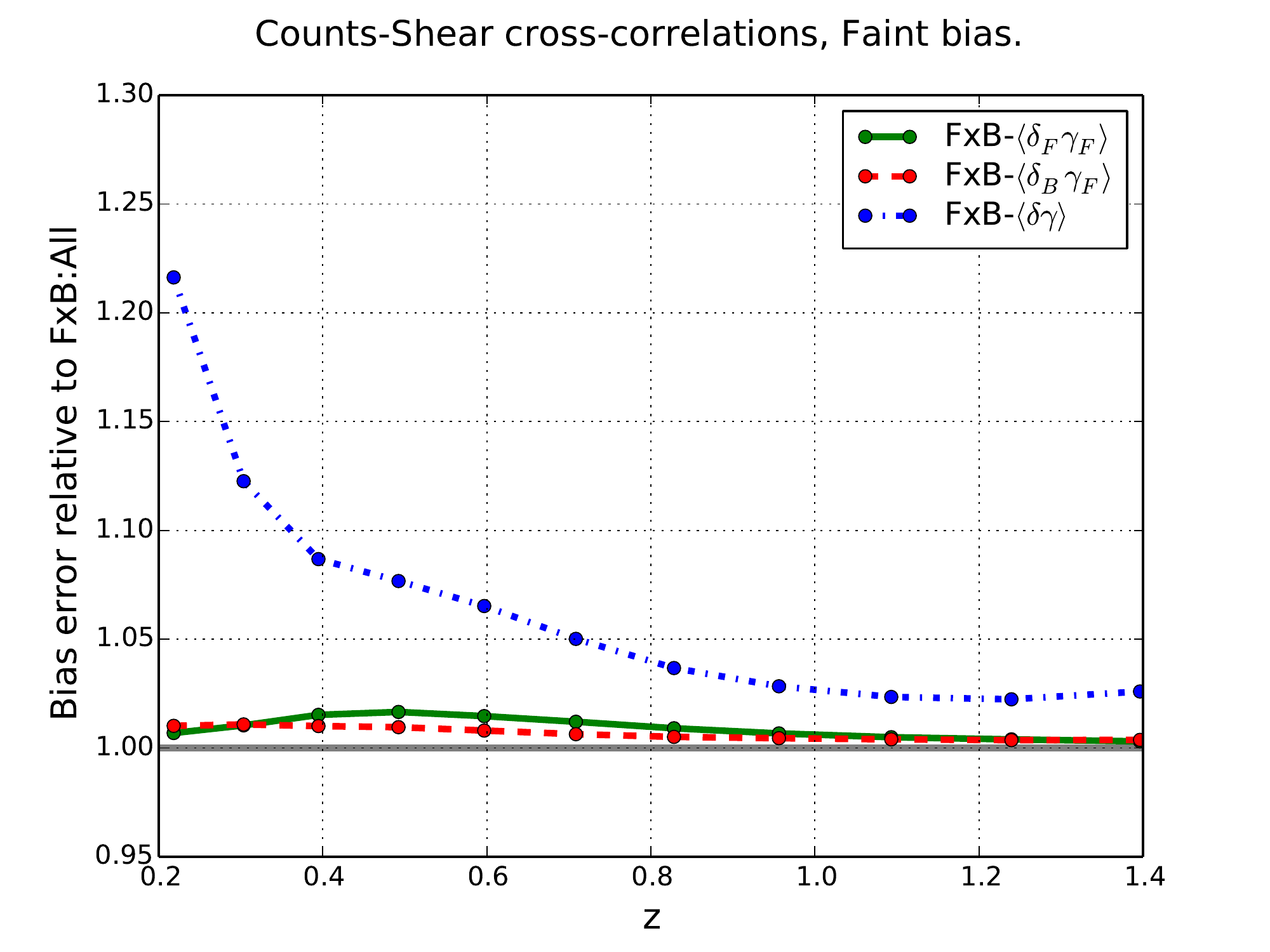}
\caption{The improvement from counts-shear cross-correlations in the Faint bias
measurements. Lines are the bias error ratios between FxB:All without some
counts-shear cross-correlations and FxB:All. In \gkone[\gktwo] all
correlations are included, except counts-shear cross-correlations within the
Faint[Bright] sample. In \gkthree\ all the counts-shear cross-correlations are
removed.}
\label{errbias_cross_gs}
\xef

In paper-II we studied how different counts-shear cross-correlations contribute
to the cosmological forecast. Overlapping surveys make it possible to
cross-correlate spectroscopic (Bright) galaxy counts with galaxy shear from a
photometric (Faint) survey. For a photometric survey, one can also
cross-correlate the counts and shear, both measured from the  photometric
survey. These two counts-shear cross-correlation types contribute about equal
to the cosmological parameter forecast(see paper-II, table 4 and 5). Removing
either counts-shear cross-correlations only lead to a small decrease in the
different figures of merit. However, the constraints drop significantly when
removing all counts-shear cross-correlations.

A contributing factor is the bias constraints from the cross-correlations. In
Fig.\ref{errbias_cross_gs} are the error ratios between FxB:All without some
counts-shear cross-correlations and FxB:All. For \gkone\ and \gktwo, we
respectively remove the cross-correlations of Faint and Bright galaxy counts
with the Faint shear. The change is small for removing either Bright or Faint
lenses, typically less than 1\%. On the other hand, when removing all
counts-shear cross-correlations (\gkthree) the bias error change significantly.
The Bright bias follow a similar trend (not shown), except counts-shear
contributing less below $z=0.4$.

This trend follows from a high correlation between the count-shear signal in
overlapping lens bins. Using the covariance for a Gaussian field (see paper-I),
the correlation coefficient between counts-shear using two galaxy populations
as lenses is

\newcommand{\cx}[1]{\left<#1\right>}

\begin{align}
r[\cx{\delta_B \gamma}, \cx{\delta_F \gamma}]
&\equiv \frac{\text{Cov}\left(\cx{\delta_B \gamma}, \cx{\delta_F \gamma}\right)}{
     \sqrt{\text{Var}\left(\cx{\delta_B \gamma}\right) 
           \text{Var}\left(\cx{\delta_F \gamma}\right)}} \\
&\approx \frac{\cx{\delta_B \delta_F}}{
     \sqrt{\cx{\delta_B \delta_B} \cx{\delta_F \delta_F}}}
\end{align}

\noindent
where $\delta_B$($\delta_F$) is the Bright(Faint) galaxy counts and $\gamma$ is
the galaxy shear overdensities. To simplify the notation, $\left< \right>$
denote a Cl cross-correlation and we suppress the redshift bin index. From
numerical tests (not shown), the terms including galaxy counts dominate.
Unlike for the galaxy counts, the RSD contribution is negligible for the
counts-shear correlations and the signal is effecively linear with the lens
bias. The strong correlation is therefore not introducing sample variance
cancellation. Instead it reduce the value of using two different galaxy
populations as lenses.

\subsection{Sample variance cancellation}
\label{subsec:volume}
\xbf
\xfigure{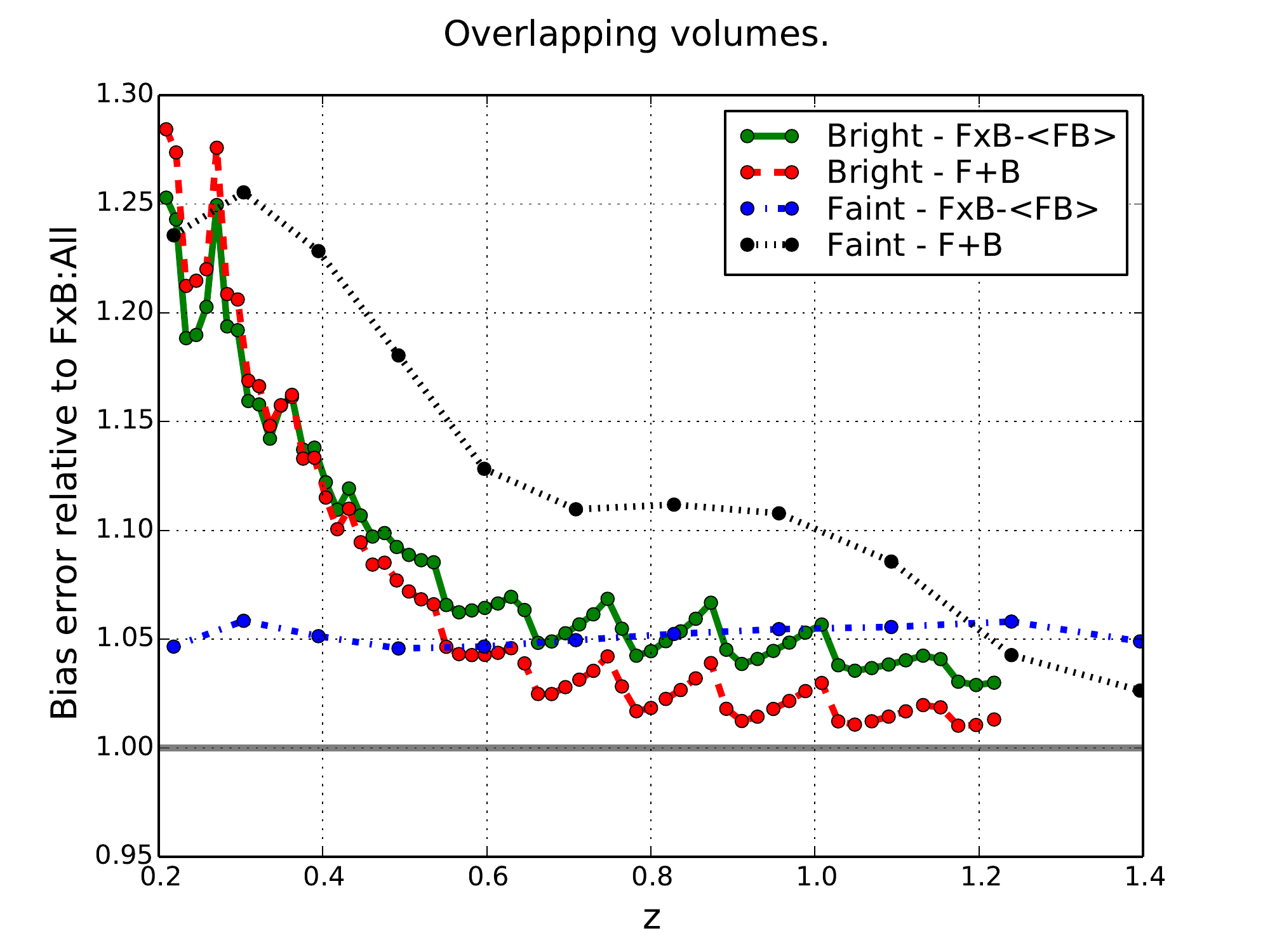}
\caption{Effect of overlapping volumes. The plots show the bias error ratios of
F+B:All and \gkfb\ with respect to FxB:All. The F+B:All combine the populations
over separate volumes, while \gkfb\ is overlapping surveys without
cross-correlating the observable. These definitions discriminate between
constraints from the covariance and additional cross-correlations.}
\label{errbias_cross_vol}
\xef

Overlapping surveys gain from both additional cross-correlations and sample
variance cancellation. Even without including the additional observables, two
overlapping surveys with free galaxy bias benefit from overlapping skies. The
overlapping surveys introduce additional covariance between the observable,
which improve the constraints when different observables respond sufficiently
different to the same variable (see \pssky). These two effects
contribute about equal to the combined forecast (paper-II, table 4 and 5).

Fig. \ref{errbias_cross_vol} distinguish between the two main effects
contribution, the additional cross-correlations and sample variance
cancellations, directly in the bias error. Both effects are included in
F+B/FxB, while \gkfb/FxB only show the contribution from sample variance
cancellation. These two effects both decrease the Faint bias error, but the
additional Bright-Faint cross-correlations contribute most. The
Faint/photometric sample has broad redshift uncertainty (see Table
\ref{tbl_assumpt}) and benefit less from RSD and BAO. The Bright-Faint
cross-correlations partly recover this information, which makes them
suitable for constraining the Faint bias (see paper-I). These 
correlations are less important for the Bright sample, where the
sample variance cancellation is most important. A single Faint redshift bin
($\dzbin = 0.07(1+z)$) overlap with multiple Bright bins ($\dzbin =
0.01(1+z)$). This introduce a covariance between the Faint bias
parameter and each of the Bright bias parameters, which then 
also indirectly correlate the Bright bias parameters.

\subsection{Bias amplitude}
\label{subsec:ampbias_abs}
The fiducial bias evolution was given in Eq. \ref{fidbias}. At redshift zero
the fiducial bias for the Bright and Faint populations equals, while the
Bright/Faint bias ratio is respectively 1.7 and 2.1 for $z=0.5$ and $z=1$.
This section introduce an additional multiplicative amplitude (relative bias
amplitude), meaning unity is the fiducial bias.

\xbf
\xfigure{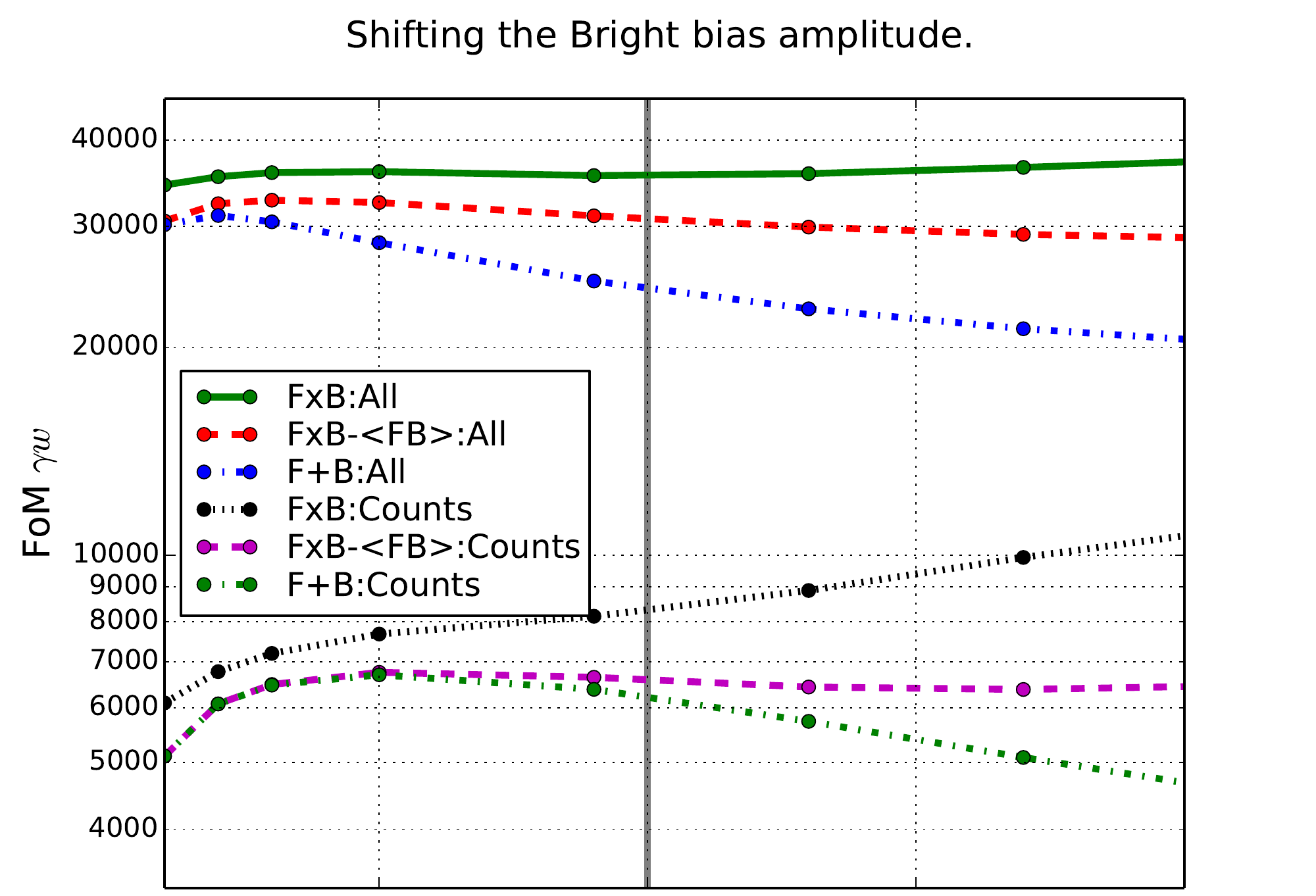}
\xfigure{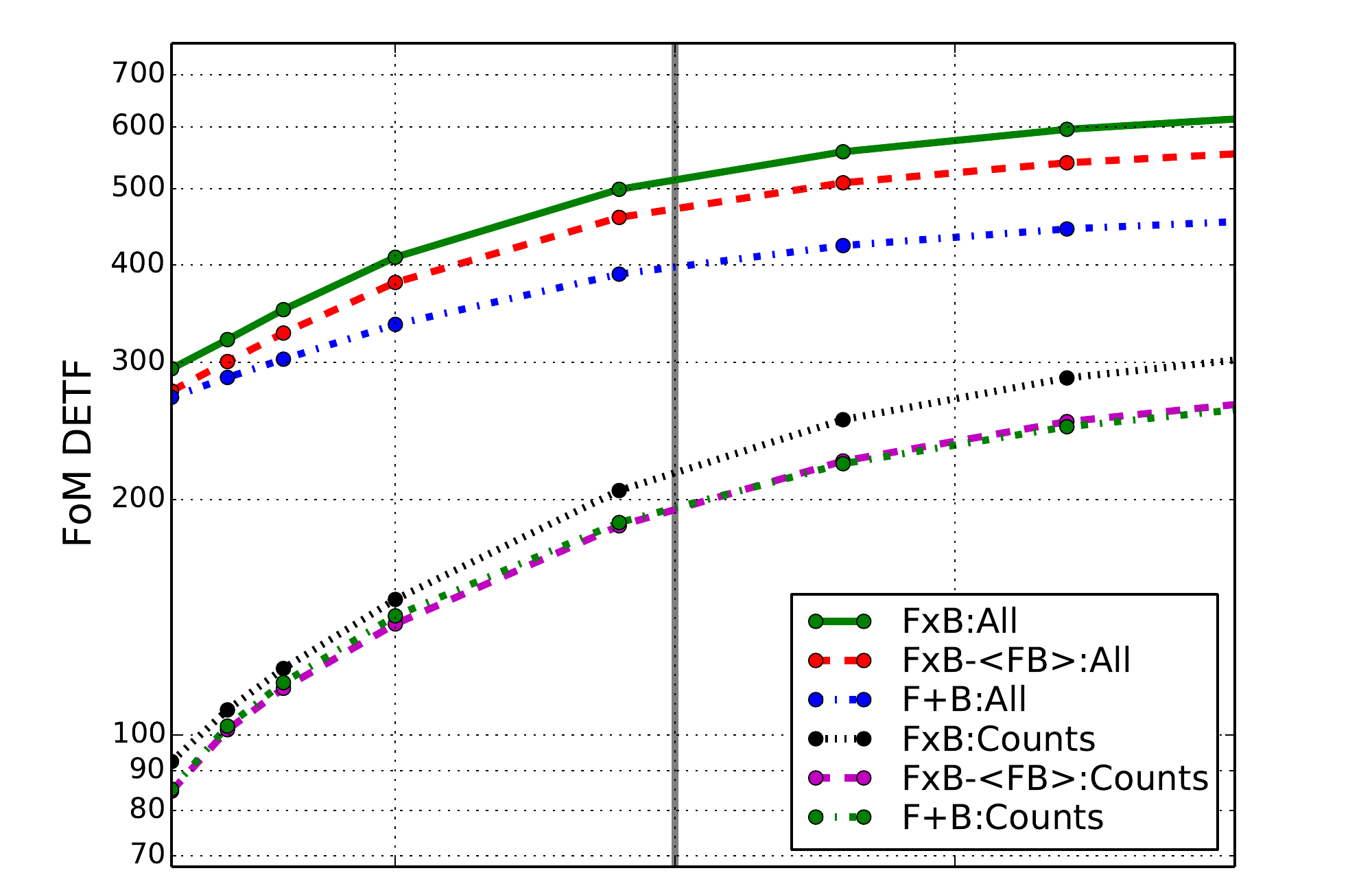}
\xfigure{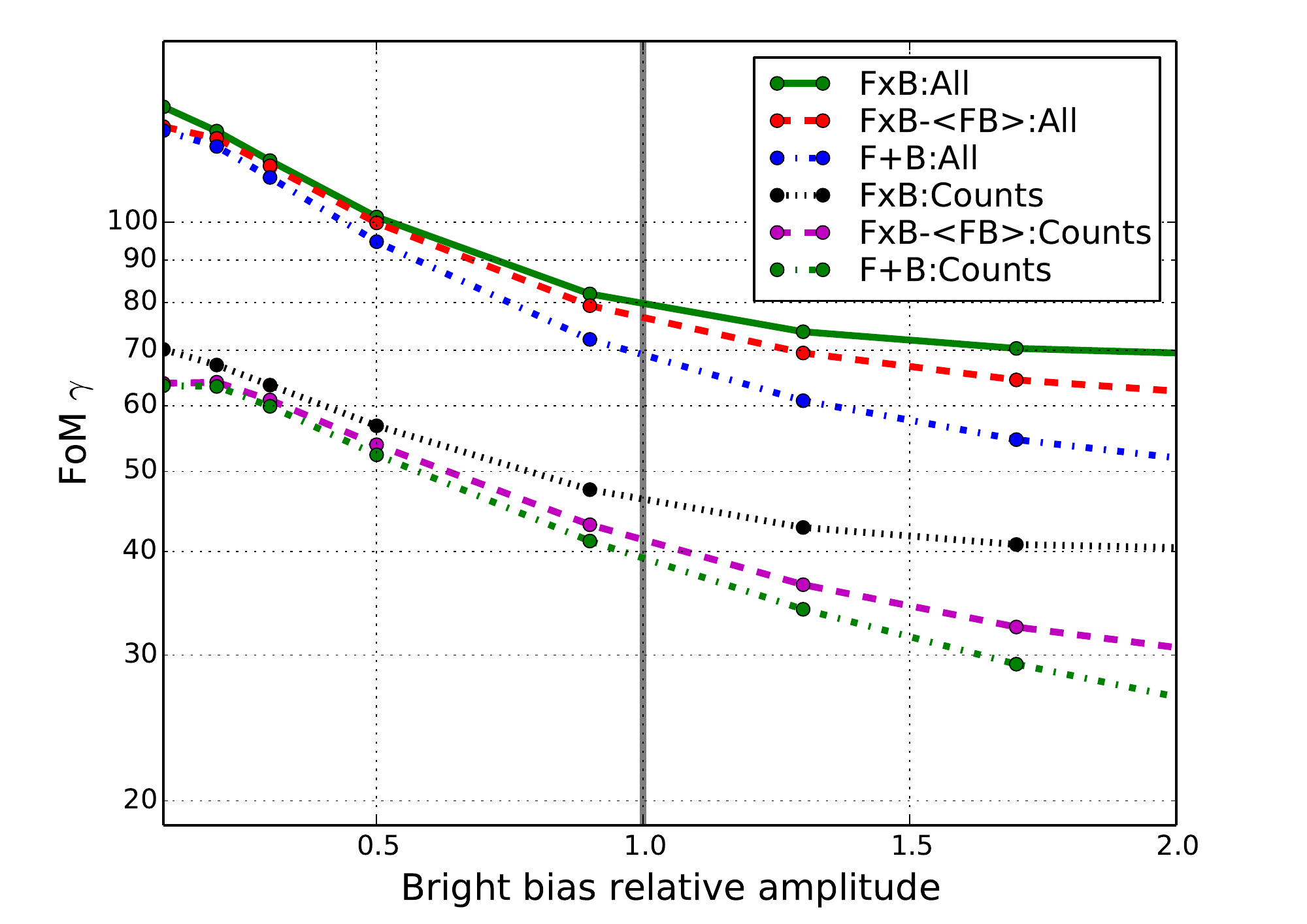}
\caption{Effect of the Bright galaxy bias amplitude on the FoMs. On the x-axis
is a multiplicative amplitude relative to the fiducial Bright bias. A vertical
line ($x=1$) marks the fiducial case. The top, middle and bottom panels
respectively show the impact for $\fomc$, $\fomdetf$ and $\fomg$.}
\label{bias_relamp_bright}
\xef

Fig. \ref{bias_relamp_bright} shows $\fomc$, $\fomdetf$ and $\fomg$ for a
shifted Bright bias amplitude. In the top panel, the different probes (lines)
gain or decline moderately with a shifted bias. $\fomc$ combine the expansion
and growth history constrains, which are affected differently by the bias
amplitude. For $\fomdetf$ (middle panel) measuring the expansion history
($w_0,w_a$), the constraints increase with the Bright bias, while the growth
constraints in $\fomg$ decline. This cause the $\fomc$ combined constraints to
be quite flat. These competing trends cause $\fomc$ to change less with the
Bright bias amplitude.

The galaxy counts shot-noise is independent of bias and depends inversely on
the surface density of galaxies, leading to a higher noise term for the galaxy
clustering
of the Bright (spectroscopic) sample. Since the auto-correlations are
proportional to $b^2$, a higher bias increase the signal and therefore reduce
the sensitivity to the shot-noise. Doubling the bias corresponds (without RSD)
to four times higher density. Without the Bright galaxy shot-noise, the trends
change completely, with the ratios being flat for a Bright bias relative
amplitude above 0.5. The trend in $\fomdetf$ is therefore caused by a higher
(lower) bias amplitude lowering (increasing) the impact of shot-noise.

Gravitational infall of galaxies towards matter overdensities along the line of
sight introduce an additional redshift, which result in the redshift space
distortions (see paper-I). The additional observed overdensity depends only on
the matter distribution and is independent of the galaxy clustering and the
galaxy bias. The galaxy bias therefore determine the relative strength of the
intrinsic clustering and the RSD signal. For a low bias amplitude the RSD
signal dominate, while the intrinsic galaxy clustering becomes more important
for a high bias. The $\gamma$ constraints ($\fomg$) are mostly from RSD and
therefore decrease with the bias amplitude. This trend also hold when not
including the Bright shot-noise.

The lines \gkfb:All [\gkfb:Counts] are FxB:All [FxB:Counts] without including
the Bright-Faint cross-correlations, but they benefit from overlapping volumes.
These lines therefore distinguish between the two sources of FoM improvements.
For FxB:All and $\fomc$ at the fiducial bias, these effects contribute about
equal. The sample variance contribute stronger when including lensing and
even when fixing the bias. We attribute this to the overlap introducing a
covariance between clustering and lensing observables, which correlates the
cosmological parameters later marginalized over. At lower Bright bias
amplitudes the sample variance cancellation contribute less for all FoMs (\gkfb
$\approx$ F+B). Without shot-noise (not shown) the overlapping samples still
benefit from sample variance cancellations, which means a sufficient density
compared to the amplitude is needed to benefit from the two tracers.

Not included are the FoMs when varying the Faint bias amplitude. The Faint sample
models a photometric survey with $\sigmaz = 0.05(1+z)$, which is more suitable
for weak lensing than galaxy clustering, since a photo-z above $\sigmaz \approx
0.005(1+z)$ erase radial information \citeind{gazta}. Compared to the Bright bias
(Fig.\ref{bias_relamp_bright}), the changes is therefore smaller when shifting
the Faint bias. However, unlike the Bright bias, the effect of shot-noise
decrease with the Faint bias (not shown). Increasing the Faint bias decrease
the bias difference between the populations and this reduce the benefit of
using two tracers.

\section{Bias priors and modeling}
\label{secbias:prandmodel}
The last section studied the bias errors and the effect of the bias amplitude.
This section looks at various aspects of the galaxy bias. The fiducial bias is
deterministic, parameterized by one parameter for each bin and include no
priors. The first subsection quantify the effect of adding priors on the bias.
In the second subsection we compare the cosmological constraints for two
different bias parameterizations. In the last subsection we add stochasticity.

\subsection{Absolute priors on bias.}
\label{subsec:abspriors}
\begin{figure*}
\zfigure{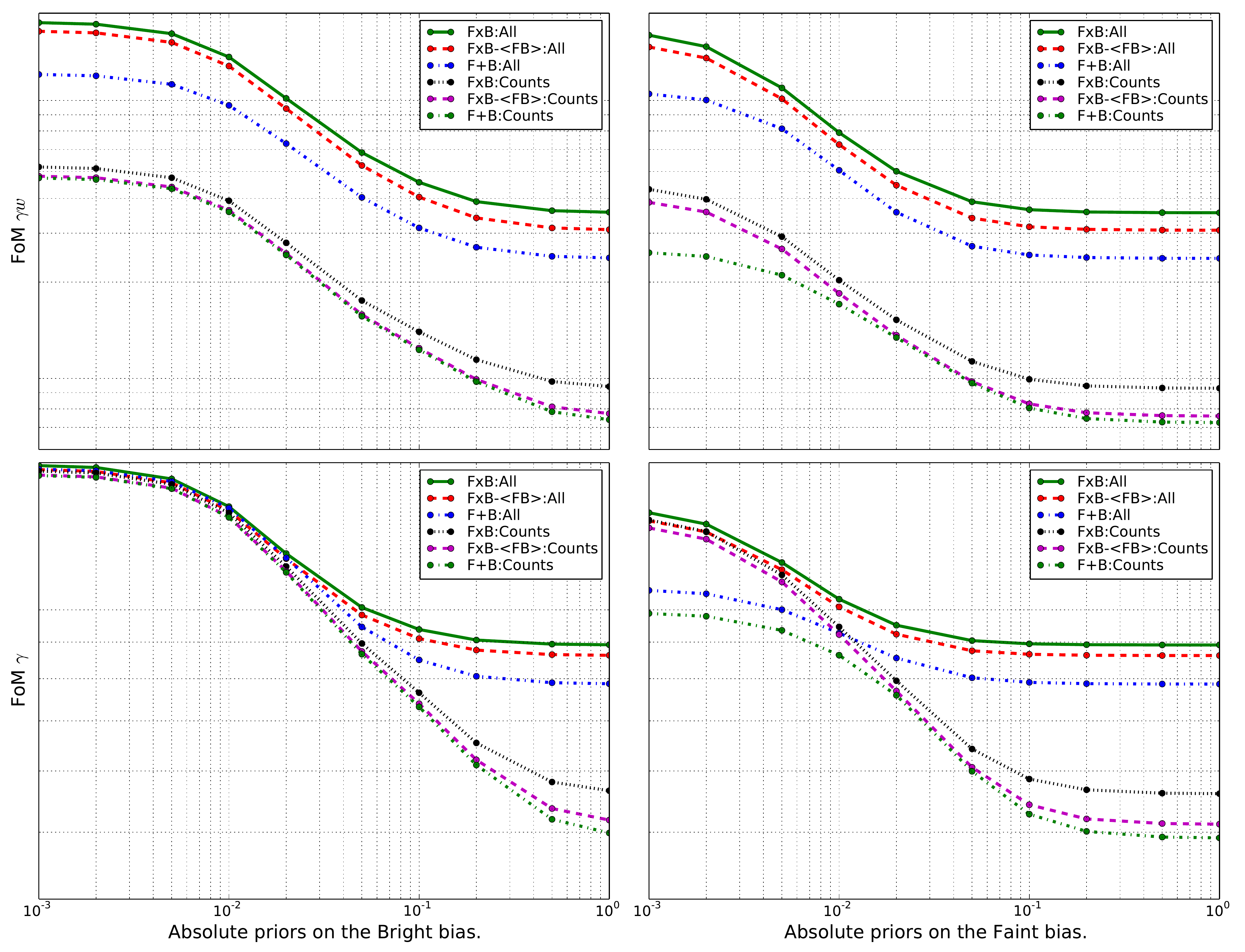}
\caption{The effect of adding priors to the bias parameters. On the x-axis is
the prior (error) added independent to each bias parameter and only for one
population. The left(right) panels add priors to the Bright(Faint) sample. The
top(bottom) panels show the effect of $\fomc$($\fomg$). Lines correspond to the
combinations FxB:All, \gkfb:All, F+B:All, FxB:Counts, \gkfb:Counts and F+B:Counts.}
\label{fombias_joint_abspriors}
\end{figure*}

This subsections study how the forecast change when adding uncorrelated priors
on the bias. Adding priors give results in between a fixed and free bias. Bias
priors from e.g. the 3pt function \citeind{3ptfunction, 3ptmes, another3pt,
yet3p, bel2015} or simulations, would usually have covariance between the parameters.
However, adding a fixed and uncorrelated prior (flat prior) on each bias
parameter simplify the model. Understanding the potential gains and required
level of accuracy is important before deciding if the improved constraints
justify the required effort. In this subsection, the bias priors is only added
to one population.

Fig. \ref{fombias_joint_abspriors} shows the forecast when adding priors on 
either the Bright or Faint bias. Here rows corresponds to the FoMs and 
columns to which population the bias is added to. The left side of
the graph is effectively fixing the bias, while on the right the priors are
weak. For FxB:Counts and F+B:Counts, the $\fomc$ and $\fomg$ forecast increase
already for very weak Bright priors (0.5). When including lensing (FxB:All and
F+B:All), stronger priors is needed, since lensing also constrains the bias.
Around absolute priors of ${10}^{-2}$ the constraints flatten and beyond this
point increasing bias priors provide no additional benefit.

How does the Bright priors affect the conclusion on overlapping versus
non-overlapping galaxy surveys? In the top,left panel stronger priors on the
Bright bias increase $\fomc$ for all probes with a factor of a few. However,
the same-sky ratios (FxB/F+B):All and (FxB/F+B):Counts are respectively close
to constant and decrease significantly. Overlapping surveys including shear
(All) are therefore equally powerful when knowing the Bright galaxy bias. For
overlapping surveys the Bright priors directly improve counts-shear
cross-correlations of Bright lenses. The Bright priors additionally lead to
stronger constraint on the Faint bias through counts-counts cross-correlations,
which again benefits the cross-correlations of Faint counts with shear. For
$\fomg$ all probes, (Fig. \ref{fombias_joint_abspriors}, bottom left panel),
approach the same asymptotic value. The growth constraints are already without
priors dominated by redshift space distortions in the Bright sample. Naturally
the Bright sample dominates the growth constraints when effectively fixing the
Bright bias and this results in no benefit from the overlap (FxB).

The right column (Fig. \ref{fombias_joint_abspriors}) add priors to the
Faint bias. In the top panel ($\fomc$), the trend for FxB:All and F+B:All is
similar to when adding priors on the Bright bias. The largest difference is for
the growth constraints ($\fomg$) in the bottom panel. For stronger priors
F+B:Counts saturates, while FxB:Counts continues improving. We have included
\gkfb, which is FxB without the Bright-Faint cross-correlations. The FxB:Counts
and \gkfb:Counts lines are close for strong Faint priors, which shows the
gain comes through sample variance cancellation (overlapping volumes) and not
the additional cross-correlations. Finally, one should remember Fig.
\ref{fombias_joint_abspriors} includes priors on only one population. When
fixing the priors, paper-II (table 5 and 6) found less benefit of overlapping
surveys.

\subsection{Comparing bias parameterizations.}
\label{subsec_biaspar}

\xbf
\xfigure{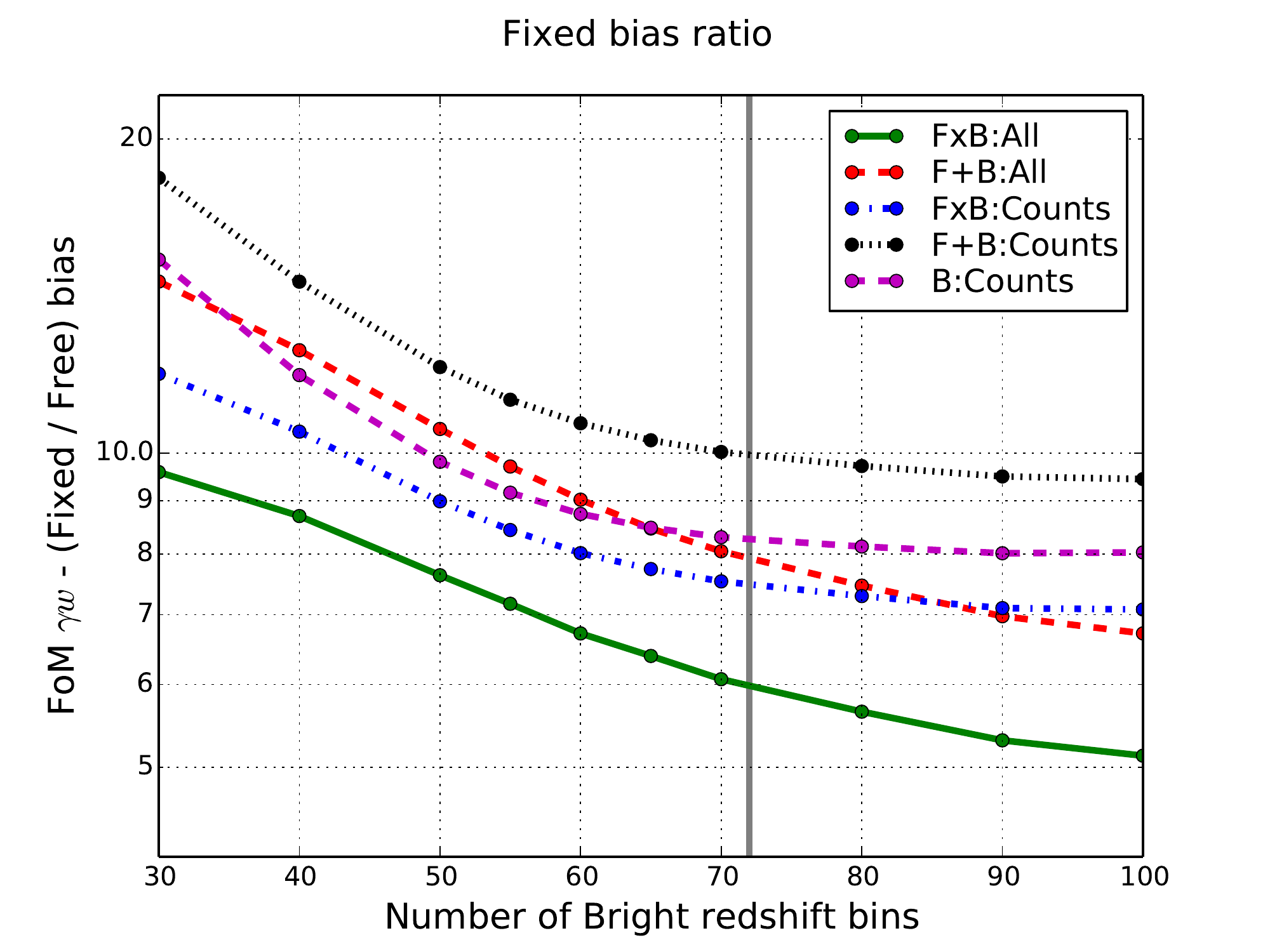}
\caption{Effect of fixing the galaxy bias for a different number of Bright
redshift bins. The $\fomc$ ratio is between a fixed and marginalized galaxy
bias, with lines corresponding to FxB:All, F+B:All, FxB:Counts, F+B:Counts and
B:Counts. On the x-axis is the number of redshift bins in the Bright sample and
the vertical line (bins = 72) mark the fiducial number of bins.}
\label{fombias_biasratio}
\xef

\xbf
\xfigure{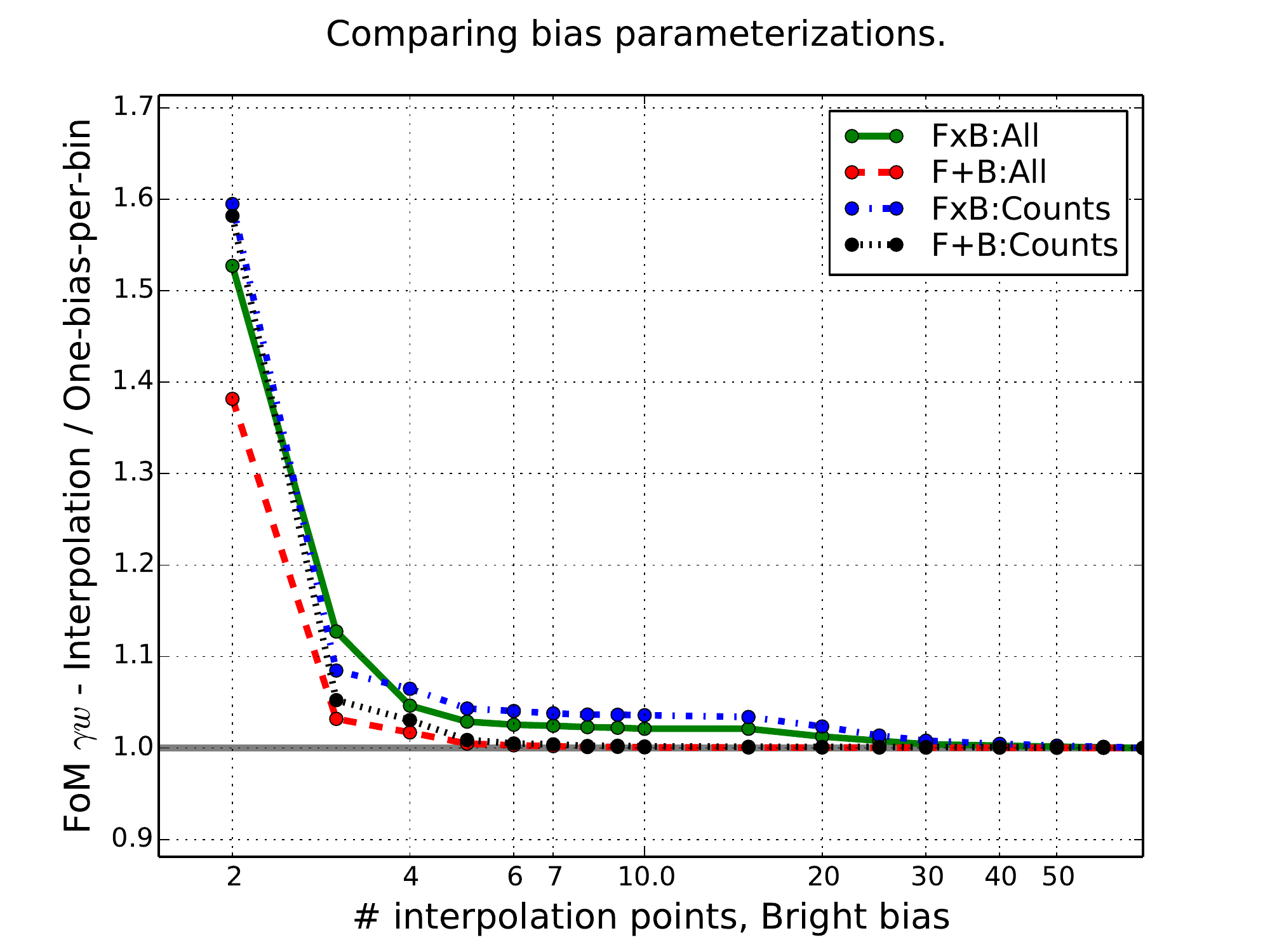}
\caption{Comparison between two different bias parameterizations. The FoM ratios are between
a linearly interpolated and the one-parameter-per-bin parameterization of the Bright
bias. The Faint bias use one-parameter-per-bin. One the x-axis is the number of
interpolation points, which scaled by $\Delta z \propto (1+z) w$ (see text). The
lines corresponds to FxB:All, F+B:All, FxB:Counts and F+B:Counts.}
\label{fombias_linbias}
\xef

Galaxy bias evolve slowly with redshift \citeind{gazta}. How many parameters
are actually needed to capture the evolution? This series of papers (paper-II,
paper-III, \pssky) parameterize the bias by one parameter in each bin and 
galaxy population. One could instead specify the bias for a few pivot
points in redshift and interpolate between them. The potential advantage is
reducing the parameter freedom and therefore increasing the statistical errors
after marginalizing over the bias. However, including sufficient freedom in the
bias parameterization will bias the cosmological constraints \citeind{clerkin}.
This subsection (and paper) focus only on the statistical errors.

Fig. \ref{fombias_biasratio} shows the ratio between Fixed/Free galaxy
bias for an increasing number of redshift bins. This partly answer the question
if the forecast becomes more or less sensitive when increasing the number of
redshift bins. Previously paper-II studied the effect of adding more Bright 
redshift bins. Here we find that increasing the number of redshift bins is 
not increasing the sensitivity to bias. On the contrary, the sensitivity to the
bias is decreasing with an increasing number of bins. The decrease is driven by
the dark energy constraints. Looking at $\fomg$, the ratios is quite flat,
while $\fom$ and $\fomdetf$ decline (not shown) and result in $\fomc$
declining.

An alternative bias parameterization is linearly interpolating between a
bias specified in pivot points \citeind{gazta}. Fig. \ref{fombias_linbias}
compare $\fomc$ when using a linear interpolated and one-bias-per-bin Bright
bias parameterization. The distance between bias pivot points scale by
$\Delta_z \propto (1+z)$, similar to the redshift binning\footnote{Paper-II
included a formula to define $N$ redshift bins in a fixed redshift range
when the redshift bin width scale by $(1+z)$. Here $z$ is the lower redshift
bin in the bin. Equivalently, the n'th pivot point redshift is

\be
z_n = -1 + (1 + z_0) {\left( \frac{1+z_{max}}{1+z_0} \right)}^{\frac{\mathlarger{n}}{N-1}}
\ee

\noindent
when locating $N$ pivot points in $[z_0, z_{max}]$. This equation also give
the n'th edge between $N-1$ redshift bins.}. Earlier, the forecast found
$\fomc$, $\fomg$, $\fom$ and $\fomdetf$ for FxB:All to become 6.0, 2.0, 3.0 and
3.4 times higher when fixing the galaxy bias (see paper-II, table 4 and 5).
When only having two Bright bias interpolation points, the $\fomc$ forecast is
40-60\% higher depending on the probe. For 3 bias points the result is
within 10\% of the one-bias-per-bin result and the two bias parameterizations
fully converge for more interpolation points. The reduction of $\fomc$ already
occurs for a few interpolation points, which is needed since the bias evolve
with redshift. We therefore conclude that using one-bias-per-bin is 
not leading to unreasonable large degradation in the statistical constraints.

\subsection{Stochasticity model}
\label{thr_stoch}
The relation between dark matter and galaxies is on large scale arguably close
to deterministic \citeind{gazta}. In these papers the fiducial bias model for
both galaxy populations are deterministic and linear in mass, meaning

\be
\delta_g = b \delta_m
\ee

\noindent
where $b$ is the galaxy bias and $\delta_g$ and $\delta_m$ are respectively the
galaxy and matter overdensities. A stochastic relation between the matter and galaxy
overdensities can be written

\be
\delta_g = b \delta_m + \epsilon
\label{delta_stoch}
\ee

\noindent
where $\epsilon$ is the random component. Any linear matter dependency in
$\epsilon$ would only lead to a redefined bias. More generally both the
bias and stochasticity could depend on higher order of the matter
fluctuation. These models are beyond the scope of this article.

The resulting power spectrum when including a stochastic component uncorrelated
with matter (Eq. \ref{delta_stoch}) is

\be
\tilde{P}(z,k) = P(z,k) + A(z,k)
\label{pk_stoch}
\ee

\noindent
where $A(z,k)$ is an additional contribution to the power spectrum $P(z,k)$. In
general this contribution can depend on galaxy population, redshift and scale.
By calibrating against simulations, one could try to measure the expected
errors, scale and redshift dependence of $A(z,k)$. One should note, such a
calibration would require several simulations to assess the impact of
cosmology. We instead focus on the direct effect on the auto-correlation

\be
C_{ii} = b^2 \left<\delta_m, \delta_m \right> + \left< \epsilon, \epsilon \right>
\ee

\noindent
for which the stochasticity introduce an additional term $\left< \epsilon,
\epsilon \right>$. We assume the stochasticity in different redshift bins are
uncorrelated, which leads to unchanged cross-correlations. Environmental
effects can introduce correlations of the stochasticity of galaxy populations,
but they are for simplicity considered independent. The stochastic component
can be modelled by

\newcommand{\cbar}{\overline{C}}
\be
\overline{C}^{xy} = C^{XY} + \delta_{XY} S
\label{corr_stoch_S}
\ee

\noindent
where $X$,$Y$ are the galaxy populations and $S$ the additional component. The
standard stochasticity function $r$ is the ratio

\be
r^2 \equiv \frac{({\overline{C}_{gm})}^2}{\cbar_{gg} \cbar_{mm}}
\label{corr_stoch_r}
\ee

\noindent
where $\cbar_{gm}$, $\cbar_{gg}$ and $\cbar_{mm}$ are respectively the
counts-matter, counts-counts and matter-matter correlations when including the
bias stochasticity. This ratio (Eq. \ref{corr_stoch_r}) relates to our
definition (Eq. \ref{corr_stoch_S}) by

\be
r^2 = \frac{C^2_{gm}}{(C_{gg} + S) C_{mm}} = {\left(1 + \frac{S}{C_{gg}}\right)}^{-1}
\label{r_expansion}
\ee

\noindent
or alternatively the relation can be written be reversed and written

\be
S = C_{gg} (1-r^{-2}).
\label{s_expansion}
\ee

\noindent
From Eq.\ref{r_expansion} and \ref{s_expansion}, values of $S$ need to be
comparable with the auto-correlations to be significant. The no-stochasticity
limit is $r = 1$ or $S = 0$. In addition to simplifying the addition of
stochasticity, the term $S$ has similar characteristics as the shot-noise.
Results on uncertainties in the stochasticity can therefore be extended to
uncertainties of the galaxy counts shot-noise.

\subsection{Bias stochasticity}
\label{fombias_stoch}
The fiducial galaxy bias in these articles is deterministic. In this subsection
we study the impact of introducing bias stochasticity (non-deterministic bias).
The last subsection (\ref{thr_stoch}) introduced a simple redshift and scale
independent parameterization of the stochasticity, deviating from the commonly
used function $r$. First we study how an increasing stochasticity reduce the
signal-to-noise and then how uncertainties in the stochasticity impacts the
forecasts.

\xbf
\xfigure{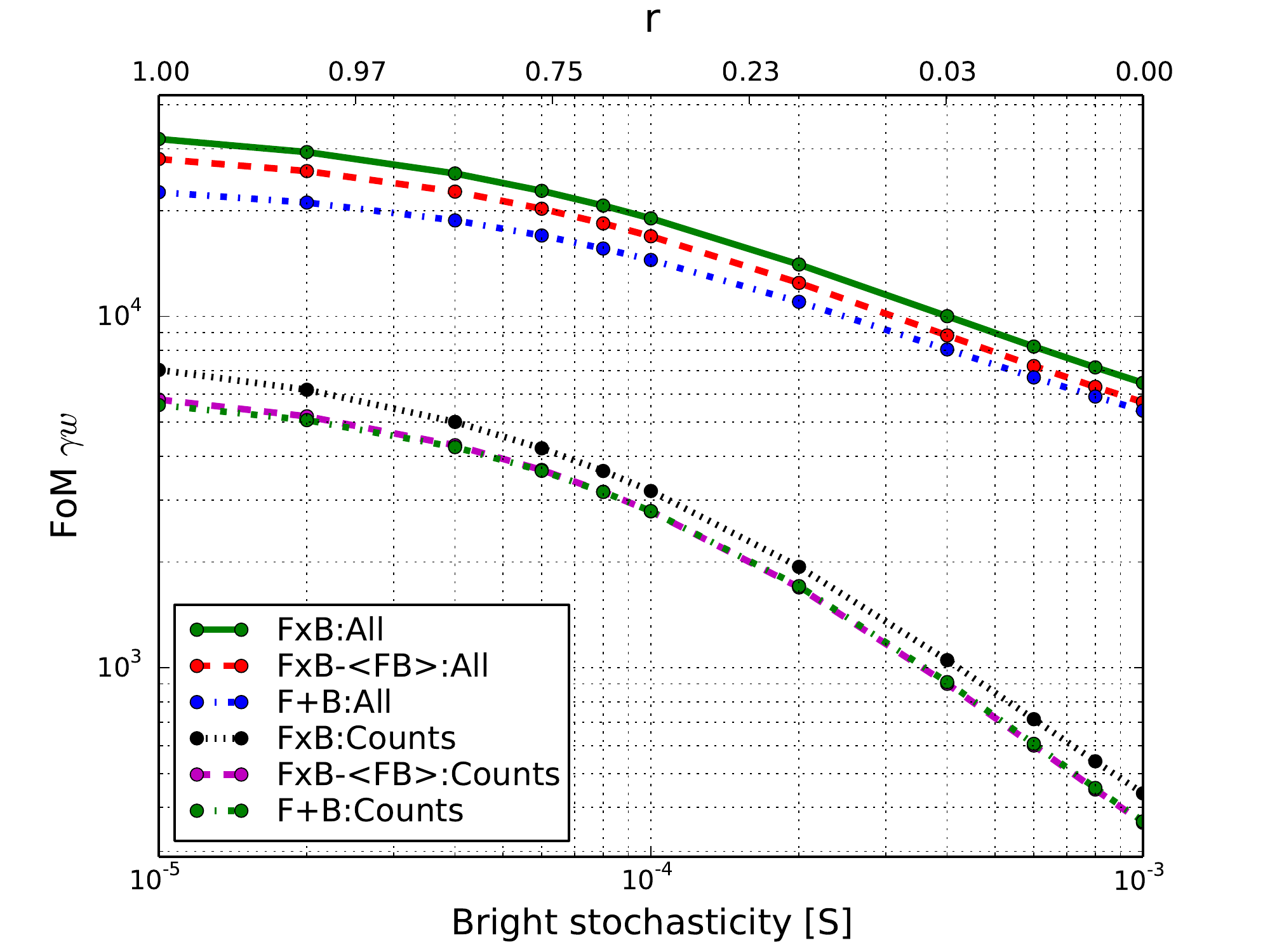}
\caption{Effect of increasing galaxy bias stochasticity. The $\fomc$ is plotted
with the x-axis being the Bright galaxy bias stochasticity parameter S. The
fiducial value is $S=0$ where the galaxy bias is deterministic and here no
extra nuisance parameters are added to describe the stochasticity. The
second/top x-axis is the stochasticity $r$ when assuming $C_{gg} =
3 \times {10}^{-4}$, which corresponds to a Bright counts-counts auto-correlation at
$z=0,l=100$.}
\label{stoch_amp}
\xef

Fig. \ref{stoch_amp} shows $\fomc$ when increasing the galaxy bias stochasticity
for the Bright population. The stochasticity variable is fixed, in other words,
the stochasticity is assumed known. Since the stochasticity is just noise, it
monotonically reduces the figure of merit. The FoMs decrease when the stochastic
component (S) comparable to the signal. For the different probes and also
$\fom$, $\fomg$ and $\fomdetf$ (not shown), the FoMs decrease steadily. The
benefit of overlapping surveys (FxB:All/F+B:All) reduce for a higher bias
stochasticity. From comparing FxB:All and F+B:All with \gkfb:All, one see the
bias stochasticity reduce the gain from sample variance cancellations. Finally,
the equivalent Faint bias stochasticity plots (not included) show only a weak
reduction. This follows from the Faint/photometric clustering contributing
weaker to the combined constraints.

\xbf
\xfigure{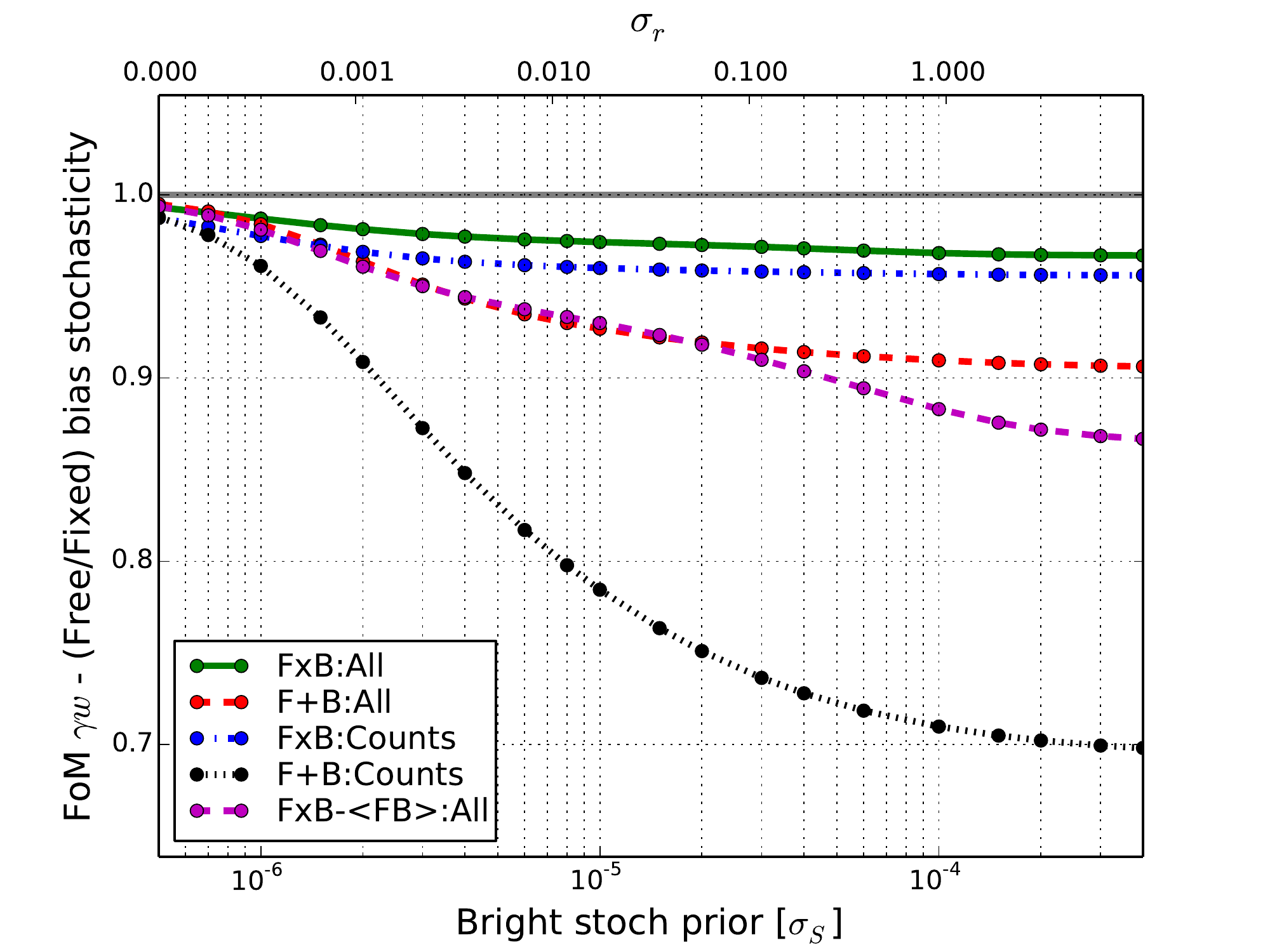}
\caption{Improvements from priors on the stochasticity. On the x-axis is prior
(error) added to all of the Bright redshift bins. The figure shows the
no-priors/priors $\fomc$ ratio.  In both populations the fiducial bias
stochasticity is zero and the Faint bias stochasticity in kept fixed. 
The second/top x-axis is the stochasticity error $\sigma_r$ when assuming $C_{gg} =
3 \times {10}^{-4}$, which corresponds to a Bright counts-counts auto-correlation at
$z=0,l=100$.}
\label{stoch_priors}
\xef

Stochasticity also reduce the parameter constraints due to additional
uncertainty in the modelling. A natural extension to the previous forecast
is to use the same parameterization as for the galaxy bias. The fiducial $S=0$
is constant in redshift, but $S$ is parameterized with one parameter per
redshift bin and population. Equivalently to the galaxy bias, simulations or
observations can give prior knowledge on the bias stochasticity. For simplicity
and similar to the bias priors, the bias stochasticity priors are assumed
uncorrelated for the different redshift bins.

Fig. \ref{stoch_priors} shows ratios between free and fixed bias stochasticity, 
when marginalizing over the uncertainty only for the Bright population.
On the x-axis is an absolute prior added to each Bright bias stochasticity
parameter. The ratio can be understood as the fraction of the FoMs recovered
from adding priors. For strong priors, on the left side of the graph, the
priors effectively fix galaxy bias and the ratio naturally approaches unity.
There is a clear difference between overlapping and non-overlapping surveys.
Without priors, the F+B:Counts reduces to 70\% of the value with known
stochasticity. The difference is 96\% of the original value when considering
FxB:Counts. Also including lensing (:All), the overlapping surveys are less
affected by the bias stochasticity uncertainties. Similar trends are found also
for $\fomg$ (not shown), but with a much smaller change.

This figure (\ref{stoch_priors}) shows that overlapping samples are less
affected by a galaxy stochasticity uncertainty. This contradicts
\citedir{kirk}, which found that marginalizing over bias stochasticity
parameters significantly reduce the same-sky benefit (from a factor of 3.9
to 1.2). Previously in \pssky, we discussed their higher absolute
gain from overlapping surveys. The main difference is \citedir{kirk} ignore the
count-shear cross-correlation with photometric (Faint) lenses. This artificially
increase the same-sky benefit, since the counts-shear signal with different
galaxy lens populations are highly correlated (see subsection
\ref{errbias_cross}). Only using the counts-shear of one galaxy population
would also increase their sensitivity the galaxy bias stochasticity.

\section{Conclusion}

This paper accompanies the previous forecast (paper-II, \pssky). These
combined a photometric (Faint) and a spectroscopic (Bright) stage-IV survey to
constrain the growth and expansion history by using a combination of weak
lensing and galaxy clustering. In particular, these papers focused on the gain
from observing the same area with both surveys. For the combined constraints,
knowing the galaxy bias would increase the figure of merit ($\fomc$)
equivalent to 3-5 times larger area. Details of the galaxy bias affected both the
same-sky benefit (i.e. FxB/F+B) and how much various effects contribute
to the constraints. This paper (paper-III) studied how the bias impacts
different aspects of the combined forecast.

Section \ref{secbias:error} focused on the bias error and 
subsection \ref{subsec:assumpt} summarized the forecast assumptions.
Then subsection \ref{subsec_biasder} introduced the bias derivative
formula, which includes the photo-z effect through the transition
matrices (see \citedir{gazta}) and also accounts for RSD and magnification.
The next subsection (\ref{errbias_fiducial}) showed the fiducial bias error and
how lensing contribute to the bias constraints. Subsection \ref{errbias_cosmo}
stressed how ignoring the covariance between the cosmology and bias 
would lead to underestimate the value of overlapping surveys for
constraining the bias. 

% Possible location to split...
Overlapping surveys both allow for additional cross-correlations and sample
variance cancellations directly from overlapping volumes (see paper-II and
\pssky). In subsection \ref{errbias_cross} we studied which counts-shear
contribute to measuring the bias. From a covariance between the signals, the
bias error only increase weakly when removing either lens population, but
strongly when removing both. Subsection \ref{subsec:volume} studied the
direct effect of overlapping volumes. The sample variance contribute most to
the Bright sample, while the Faint sample benefit most from the additional
cross-correlations.

Section \ref{secbias:prandmodel} studied the effect of adding priors, bias
parameterization and modelling of the bias stochasticity. Subsection
\ref{subsec:abspriors} added uncorrelated priors in each bins on either the
Bright or Faint bias. When adding bias priors, they directly affect the galaxy
population and also the other galaxy population through the cross-correlations.
Completely fixing the galaxy bias reduce the same-sky benefit (see paper-II).
If adding priors to either population, the $\fomc$ (FxB/F+B):All same-sky ratio
only change weakly. For $\fomg$ the Bright sample would completely dominate
$\fomg$ if fixing the Bright bias. Subsection \ref{subsec_biaspar} tested the
sensitivity to the bias parameterization, which fiducially use
one-bias-per-bin. Including more Bright redshift bins reduced the sensitivity
to the bias. We further compared to a bias interpolated between $n$ redshift
pivot points. For $n=3$ the $\fomc$ results agreed within 10\% and the
one-bias-per-bin is therefore a reasonable bias parameterization.

Appendix \ref{app:amplitude} studied how the fiducial bias change
the forecast. Changing the bias amplitude affects both the combined constraints
and the relative performance of various probes (clustering, RSD, WL). In
subsection \ref{subsec:ampbias_abs} the dark energy ($\fomdetf$) and growth
($\fomg$) constrains respectively increase and decrease when increasing
the Bright galaxy bias amplitude. Increasing the Bright/spectroscopic bias
reduces the shot-noise and therefore improving the dark energy constraints.
The growth constraints reduced since the relative contribution of RSD, which is
important to measure $\gamma$, decrease for a higher bias. In subsection
\ref{subsec:rsd} and \ref{subsec:bao} we studied respectively respectively the
relative contribution of RSD and BAO. Last, subsection \ref{subsec:wlamp}
investigated how the Faint bias amplitude impacts shear and magnification. A higher
Faint bias increased the importance of WL for dark energy constraints ($\fom$)
and reduces the impact of magnification. Appendix \ref{app:errsurvey} showed
how bias errors depend on photo-z and density.

The fiducial bias is deterministic. Subsection \ref{thr_stoch} introduced
a simple stochastic galaxy bias model and related the definition to the
commonly used correlation coefficient $r$ (see also \citedir{stoch_bias_var}).
Increasing the Bright bias stochasticity reduced the same-sky benefit
(FxB:All/F+B:All), with the sample variance cancellation being more affected
than the Bright-Faint cross-correlations. If marginalizing over an uncertainty
in the bias stochasticity (see subsection \ref{fombias_stoch}), the overlapping
photometric and spectroscopic (FxB) are less affected (than F+B). Thus
overlapping surveys not only provide a better figure of merit (equivalent to
50\% larger area) than separate surveys, but they are more robust to systematic
errors, such as bias stochasticity or uncertainties in the bias evolution.

\section*{Acknowledgements}
We would like to thank the group within the DESI community looking at
overlapping surveys. M.E. wish to thank Ofer Lahav, Henk Hoekstra and Martin
Crocce in his thesis examination panel, where these results were discussed. 
M.E. further thank Jacobo Asorey for discussions. Funding for this project was
partially provided by the Spanish Ministerio de Ciencia e Innovacion (MICINN),
project AYA2009-13936 and AYA2012-39559, Consolider-Ingenio CSD2007- 00060,
European Commission Marie Curie Initial Training Network CosmoComp
(PITN-GA-2009-238356) and research project 2009-SGR-1398 from Generalitat de
Catalunya. M.E. was supported by a FI grant from Generalitat de Catalunya. M.E.
also acknowledge support from the European Research Council under FP7 grant
number 279396.

\appendix
\section{Bias amplitude}
\label{app:amplitude}
Subsection \ref{subsec:ampbias_abs} studied the absolute effect of 
shifting the bias amplitude. This appendix focus on the relative
effect of RSD (subsection \ref{subsec:rsd}), BAO (subsection 
\ref{subsec:bao}) and WL (subsection \ref{subsec:wl}).

\subsection{Redshift Space Distortions}
\label{subsec:rsd}
\xbf
\xfigure{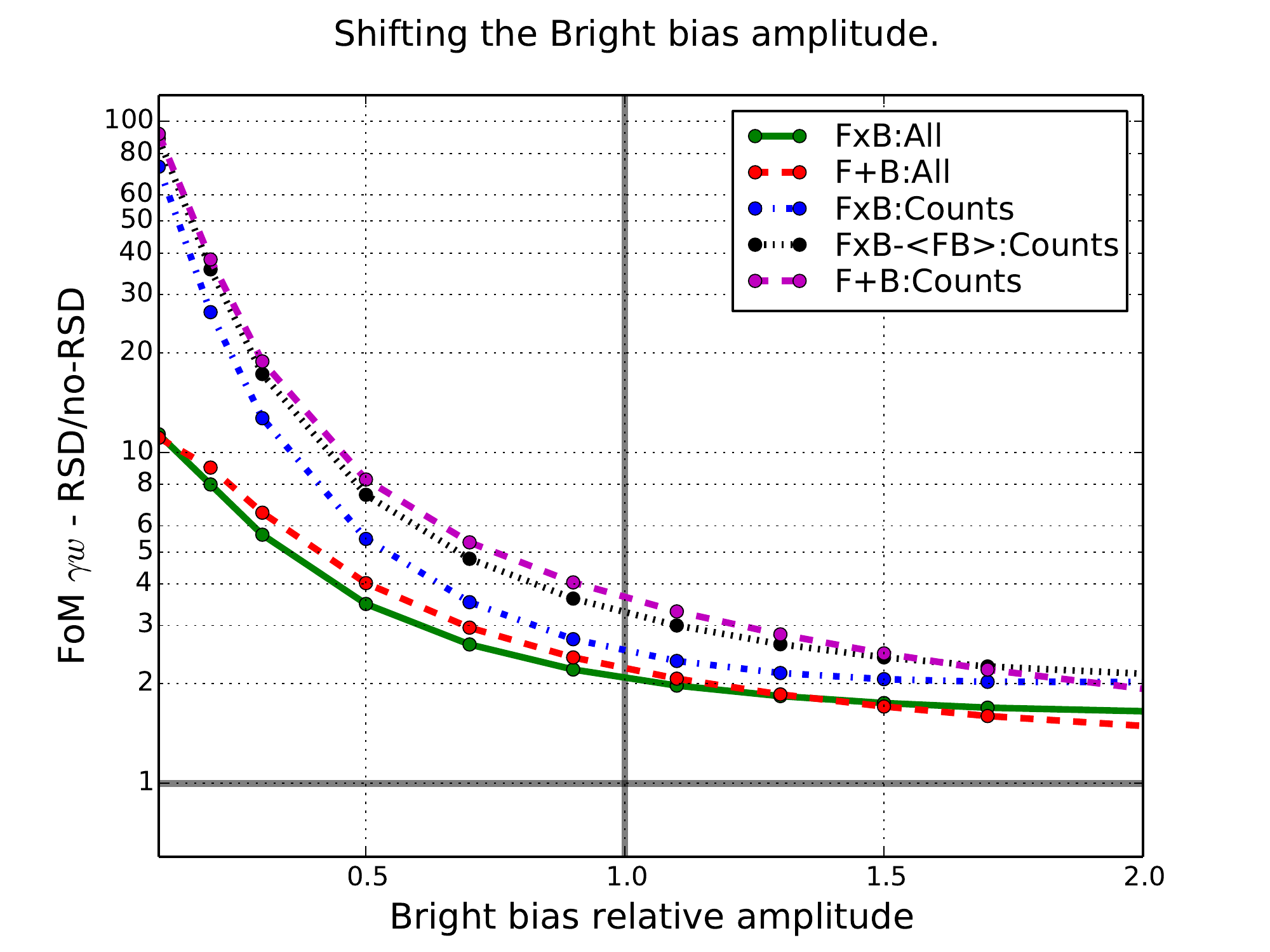}
%\xfigure{prod_fomgr49_fomg.pdf}
\caption{The effect of RSD when scaling the Bright bias amplitude. On the
x-axis is the Bright bias amplitude relative to the fiducial value. The figure
shows the $\fomc$ redshift/real space ratio. A ratio above unity means RSD 
increase the FoM. The lines corresponds to FxB:All, F+B:All, FxB:Counts,
\gkfb:Counts\ and F+B:Counts.}
\label{relbiasrsd}
\xef

Fig. \ref{relbiasrsd} shows the $\fomc$ redshift/real space forecast ratios,
varying the Bright bias amplitude. As in the previous subsections, we do not
show the ratio when varying the Faint bias. The benefit of RSD decrease for a
higher Bright bias, which confirms the importance of RSD in Fig.
\ref{bias_relamp_bright}. As expected, the largest contribution comes when
only including Counts. For low Bright biases the RSD is the main contribution
to the signal. The ratios looks dramatically high (60-100) when only including
galaxy counts. While the effect of RSD is more important for a low bias, the
RSD increasing the auto-correlation amplitude and therefore lower the impact of
shot-noise. Without the Bright galaxy shot-noise (not shown), the RSD/no-RSD
at low Bright bias drops. For a relative bias of 0.1, the ratio is 10(17) for
FxB:Counts(F+B:Counts).

For $\fomc$ the RSD/no-RSD ratio is above unity and RSD improve the
constraints. This conclusion depends on the FoM and the galaxy density. The RSD
contribute strongly to measuring $\gamma$, but less for the dark energy
constraints. For $\fomdetf$ and sufficient density, the RSD signal decrease the
constraint (not shown). While the additional RSD signal is good to measure
$\gamma$, the RSD suppress cross-correlations between close redshift bins
redshift (see paper-I). As seen in paper-II, the radial information in these
cross-correlations contribute strongly to the dark energy constraints.  This
suppression therefore lower the constraints when including RSD.

At large Bright bias amplitudes the RSD/no-RSD ratio changes differently for
overlapping (FxB) and non-overlapping (F+B) surveys. The similarity at low bias
is due to a low density and disappear when removing the shot-noise (not shown).
Because FxB include the Bright-Faint cross-correlations, one expect a smaller
dependence on the RSD contribution in the Bright sample and smaller reduction in
the RSD/no-RSD ratio. Notice also how \gkfb:Counts is only slightly below
FxB:Counts. While this trend is not the strongest, RSD is important for sample
variance cancellations. Otherwise, the galaxy clustering of both populations is
directly proportional to the matter overdensities. That would remove the benefit
of using two tracers. This effects also contribute to the FxB ratios flattening
for a high Bright bias.

\subsection{Baryon Acoustic Oscillations (BAO)}
\label{subsec:bao}
\xbf
\xfigure{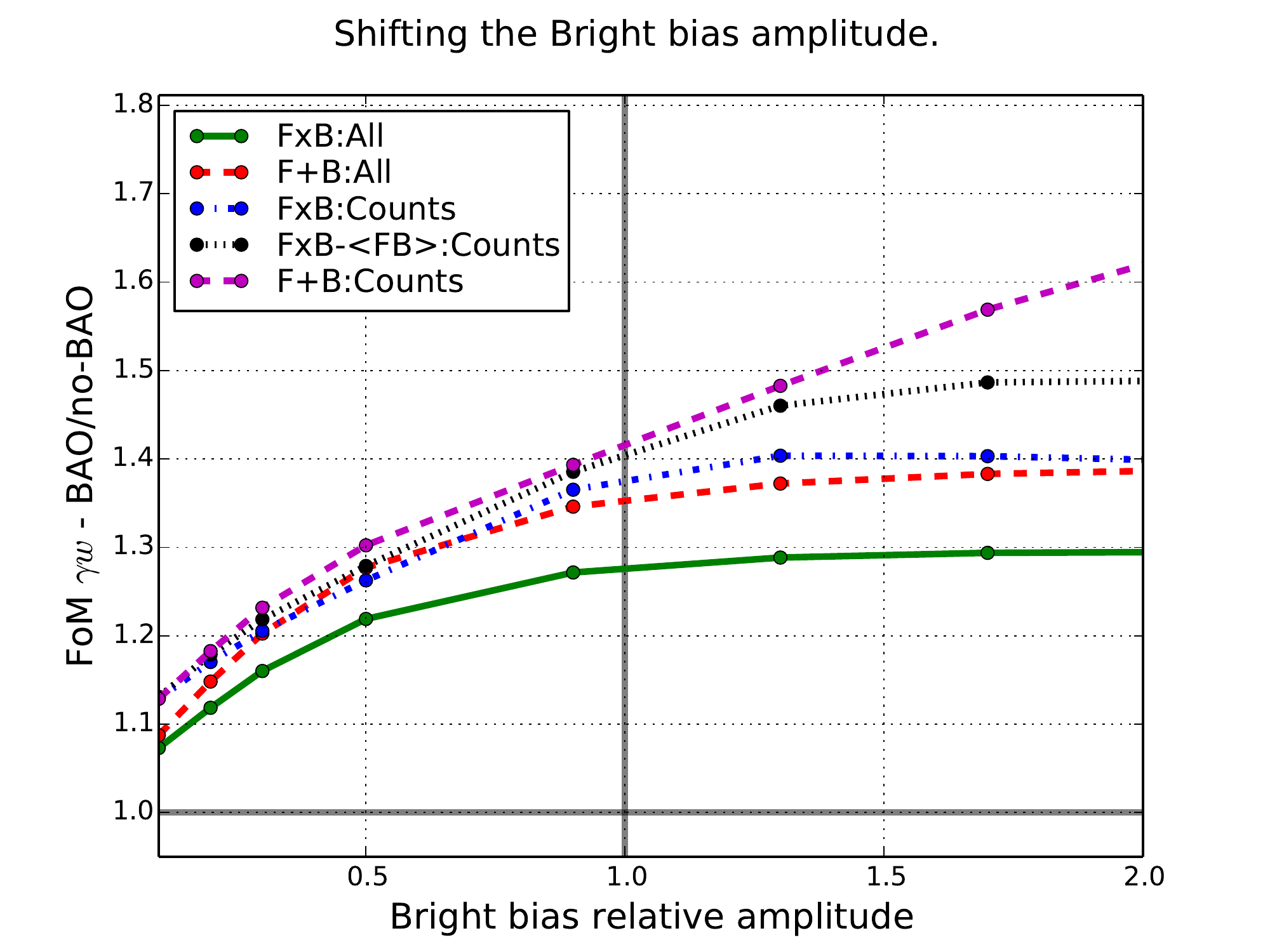}
\caption{The effect of BAO for different bias amplitudes. The $\fomc$ ratios are
between including or not BAO wiggles in the Eisenstein-Hu power spectrum. On
the x-axis is a multiplicative bias amplitude relative to the fiducial
Bright(Faint) bias. The lines corresponds to FxB:All, F+B:All, FxB:Counts,
\gkfb:Counts\ and F+B:Counts.}
\label{relbiasbao}
\xef

The BAO scale of 150 Mpc is a characteristic scale with a higher probability of
finding a galaxy pair. In the 2D and 3D configuration space correlation
function, the BAO is a significant peak, where the position is largely
independent of the galaxy bias \citedir{eisenstein2005}. Instead of only constraining
cosmology from the peak position, these papers include the BAO signal through the 2pt
correlation function. The advantage is utilizing the full power spectrum, but
it does require modelling the galaxy bias. To estimate the importance of BAO,
the Eisenstein-Hu (EH) model can predict the dark matter power spectrum with
and without BAO wiggles. We check how BAO impacts the constraints by
calculating the FoMs with and without the BAO wiggles.

Fig. \ref{relbiasbao} shows the BAO/no-BAO $\fomc$ forecast ratios when varying
the Bright bias amplitude. The BAO peak is effective in measuring the expansion,
but not the growth. Including BAO improve $\fomc$($\fomdetf$) with 30\% (23\%),
while $\fomg$ decrease with 1\%. Since BAO has largest effect on the dark
energy constraints, trends in this figure can be compared to $\fomdetf$.
Similar to the absolute $\fomdetf$ constraints, the BAO/no-BAO ratio trend
change with the Bright galaxy density. Without shot-noise, the BAO/no-BAO
ratios above a relative bias amplitude of 0.5 is close to flat (not shown). A
higher bias is therefore advantageous for reducing the shot-noise, which is why
Red Luminous Galaxies (LRGs) are often targeted for BAO studies
\citeind{eisenstein2005,cabreplot}.

\label{subsec:wlamp}
\subsection{Galaxy shear and Magnification.}
\label{subsec:wl}

Weak lensing also depend on the galaxy bias. For the signal, the counts-shear
cross-correlations and magnification are nearly proportional to the foreground
bias amplitude. Magnification, which change the observed number counts through
lensing (see paper-I), is included by default. While the measurement noise is
independent of bias, the dominant (cosmic variance) contribution also scale
linear with bias. In this subsection, we first study the effect of both weak
lensing signals when shifting the Faint bias amplitude. Next we
focus on the contribution from magnification.

\xbf
\xfigure{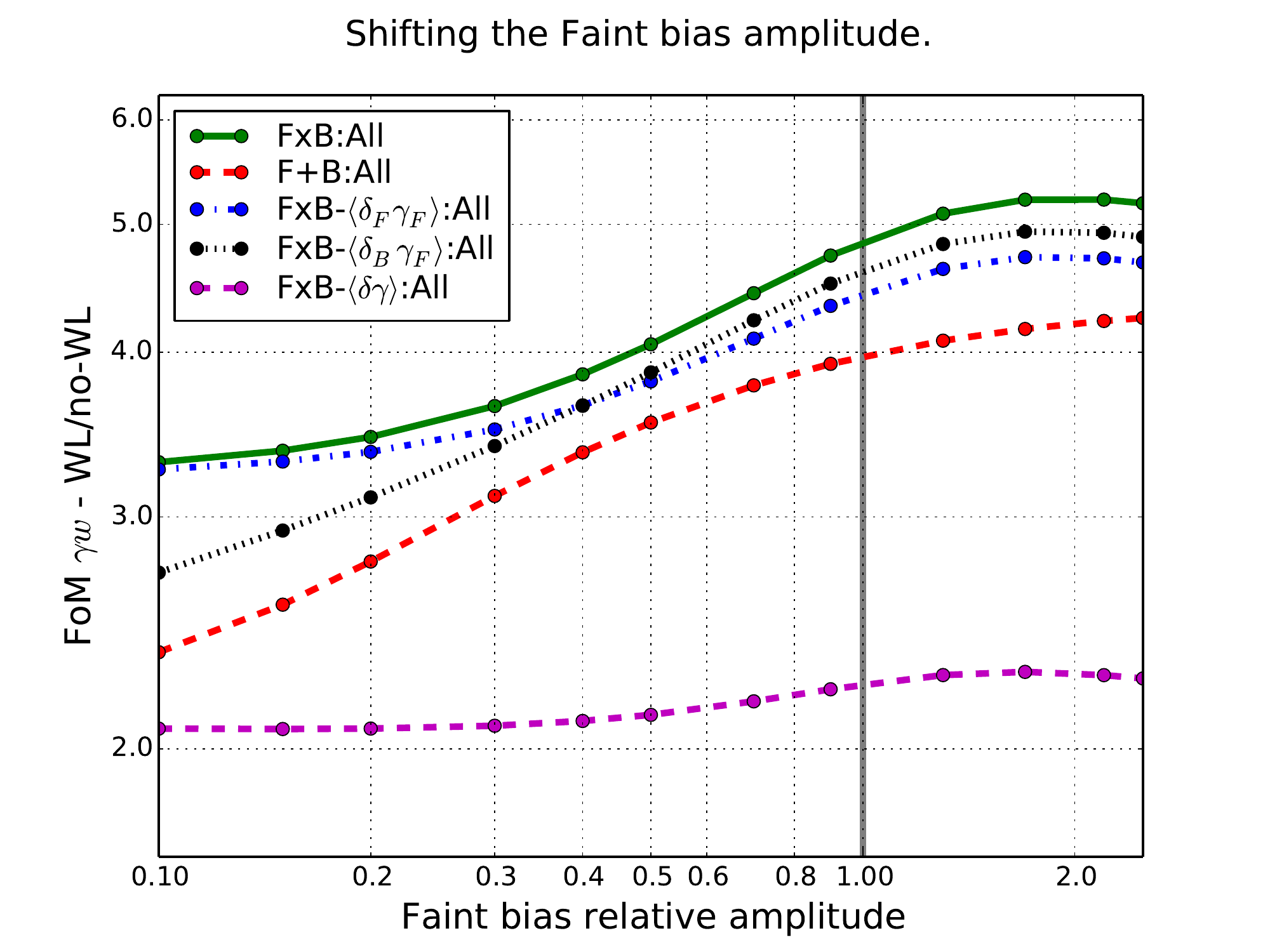}
\xfigure{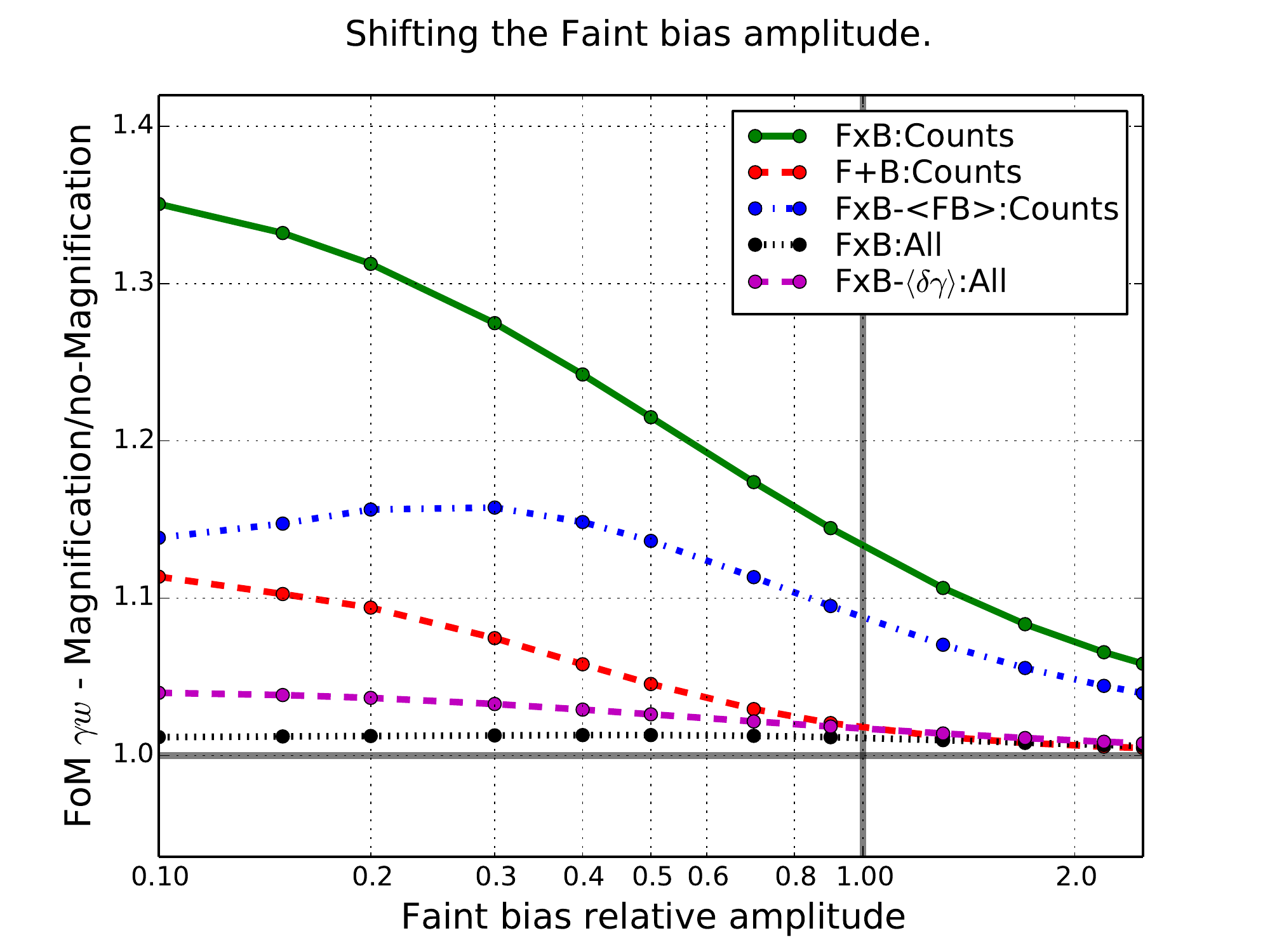}
\caption{The effect of the Faint bias amplitudes on the weak lensing
contribution. On the x-axis is a multiplicative amplitude relative to the
fiducial Faint bias. Ratios in the top panel show the WL/no-WL ratios, where
WL is the fiducial forecast and no-WL remove all WL signals (counts-shear,
shear-shear and magnification). In the bottom panel the no-Magnification
forecast set the magnification slopes (see paper-I) to zero.}
\label{relbiasWL}
\xef

Fig. \ref{relbiasWL} (top panel) shows the WL/no-WL $\fomc$ ratio. In FxB:All
the signal include the shear-shear and counts-shear cross-correlations with
both populations as lenses. This configuration naturally has the largest effect
of weak lensing. The next two lines are \gktwo(\gkone), which is FxB:All, but
removing the counts-shear correlations with Bright(Faint) lenses. In paper-II
the Faint lenses contributed slightly stronger to the forecast (see also
subsection \ref{errbias_cross}). The relative strength of different lens
populations depend on the bias amplitudes, which results in the lines crossing.
For non-overlapping surveys (F+B:All), one can only measure counts-shear with
Faint lenses and result in F+B:All and \gktwo\ following a similar trend.
Finally, the line \gkthree\ is including no counts-shear correlations and
the trend is therefore flat.

The bottom panel Fig. (\ref{relbiasWL}) shows the magnification
/no-magnification $\fomc$ ratio. Magnification affect number counts in two
ways. Foreground foreground matter over and under densities respectively
magnify and de-magnified background galaxy fluxes and areas. The magnified
fluxes can enter/leave a magnified limited sample, changing the total number of
galaxies, while the magnified area modify the galaxy density. Here the
magnification slope (magnification strength) equals the fiducial configuration
in paper-II (see paper-II, Fig. 1).

The effect of number counts magnification is small compared to WL, RSD or BAO.
For FxB:All (flattest ratio) magnification only increase $\fomc$ with $\sim
1\%$, while the effect is 15\% for FxB:Counts. This discussion therefore focus
on the number counts results (Counts). In \gkfb\ the surveys overlap, which give
sample variance cancellations, but does not include the Bright-Faint
cross-correlations. This reduce the bias constraints and therefore the
magnification constrains. The non-overlapping surveys (F+B) does not have
sample variance cancellations from overlapping volumes and the ratio drops
further. Finally, could the low magnification effect be caused by the
counts-shear and magnification signal being proportional? The line \gkthree\ is
FxB:All, but removing all counts-shear cross-correlations or equivalently
FxB:Counts adding shear-shear. The effect of magnification is also low when
including the shear-shear lensing observable.

A lower Faint bias amplitude improve the forecast. One can understand this from
the signal-to-noise. The variance of the magnification signal when assuming
Gaussian fluctuations (see paper-II) is

\begin{align}
\Delta^2 C_{\delta_i \delta_j} 
&= N^{-1}(l)[C_{\delta_i \delta_i} C_{\delta_j \delta_j} + C^2_{\delta_i \delta_j}] \\
&\approx N^{-1}(l) C_{\delta_i \delta_i} C_{\delta_j \delta_j}
\label{fombias_error}
\end{align}

\noindent
where $i$ and $j$ are the two redshift bins. For well separated bins in photo-z
space, the auto-correlation terms dominate, which least to the approximation in
the second line. If ignoring RSD, magnification-magnification correlations and
the shot-noise, the magnification signal-to-noise is

\be
\sn
\approx \left|\frac{\alpha_j}{b_j} \right| \sqrt{N(l)}
\frac{C_{m_i \gamma_j}}{\sqrt{C_{m_i m_i} C_{m_j m_j}}}
\label{fombias_magnopt}
\ee

\noindent
where $\alpha_j$ and $b_j$ are respectively the magnification slope and the
galaxy bias in bin $j$. The $C_{m_i m_i}$ and $C_{m_j m_j}$ terms are
respectively the matter auto-correlations in bin $i$ and $j$. This
approximation is invalid for low Faint bias amplitudes. Eq.
\ref{fombias_magnopt} shows two important criteria for magnification.  Higher
magnification slope increase the signal, while a lower bias of the background
population decrease the errors. This criteria can be useful if focusing solely
on magnification. To optimize the combined constraints, one should however
focus on RSD, BAO and WL, since they contribute stronger.

\section{Photometric redshifts and galaxy densities}
\label{app:errsurvey}
\subsection{Photometric redshifts}
\xbf
\xfigure{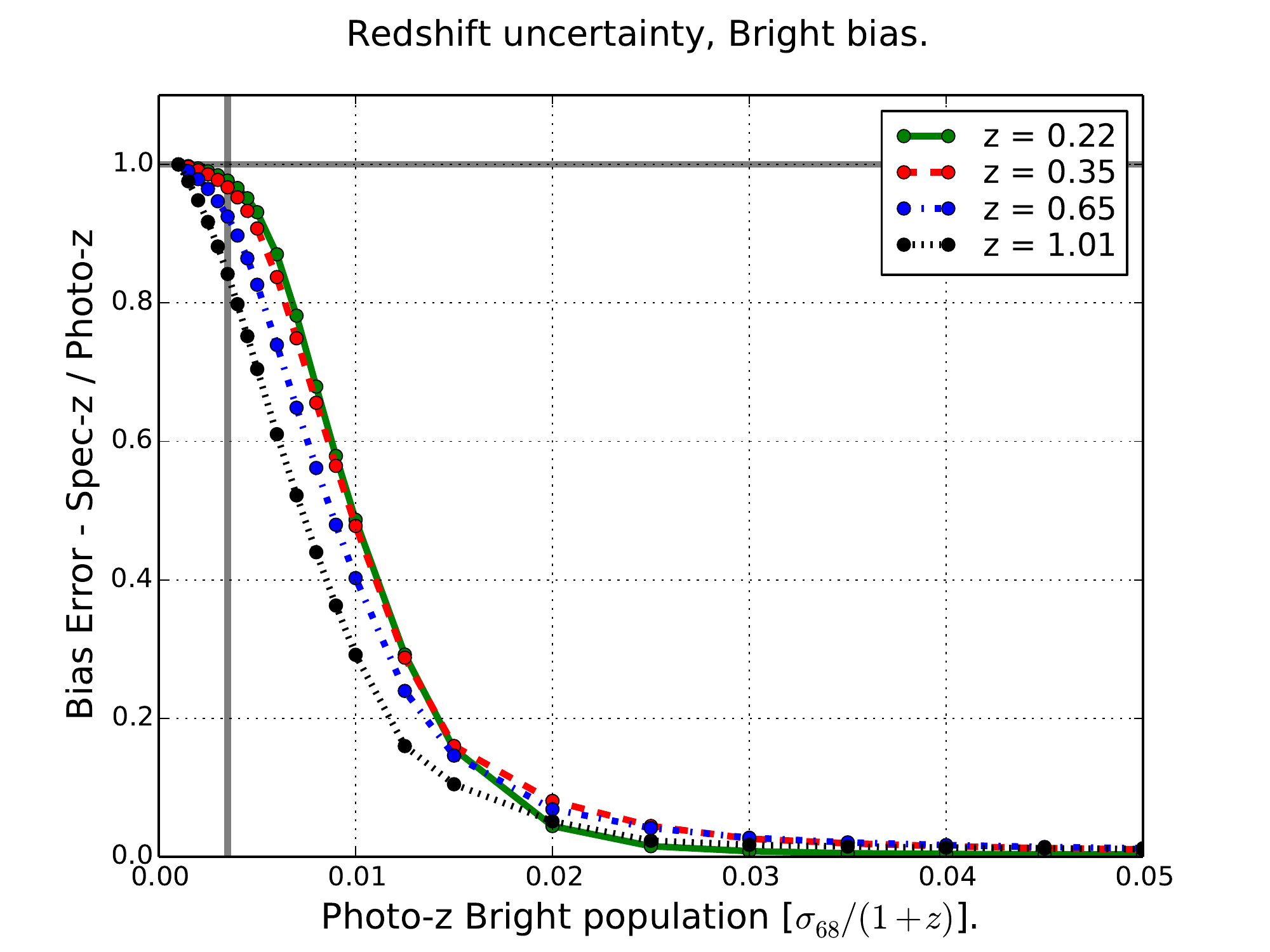}
\xfigure{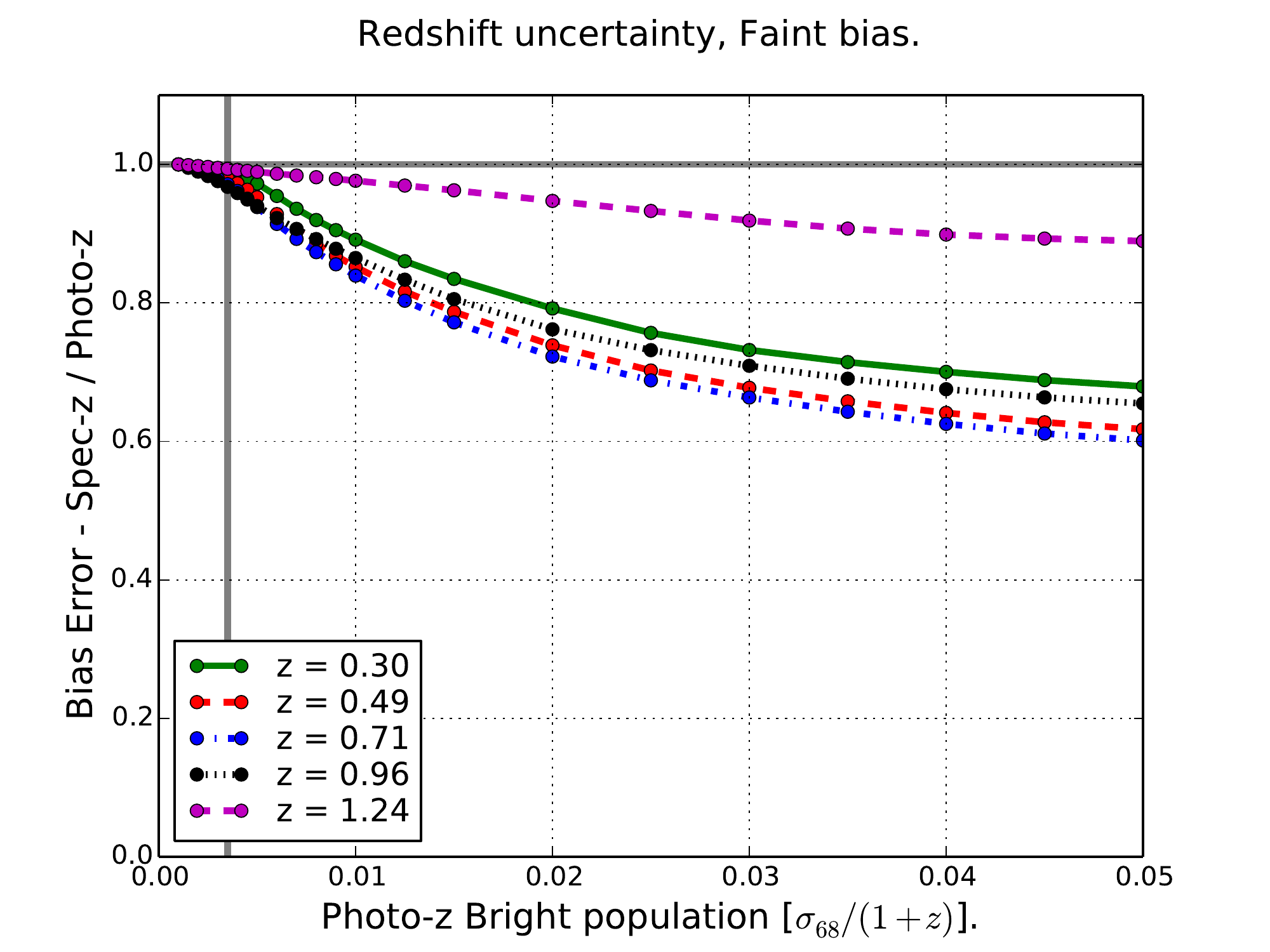}
\caption{The impact of the Bright photo-z on the bias error. Lines show the
spec-z/photo-z bias error ratios. On the x-axis is the Gaussian photo-z (units
of (1+z)) of the Bright sample. The top(bottom) panel show the ratios for the
Bright(Faint) bias.}
\label{errbiasphotoz}
\xef

The bias measurements are affected by redshift uncertainties. Our two fiducial
galaxy samples are: a photometric (Faint) and spectroscopic (Bright) with
respectively $0.05(1+z)$ and $0.001(1+z)$ Gaussian redshift uncertainties.
Since the Faint photo-z is broad, the Faint sample is analyzed in broad
redshift bins. Improving the Faint photo-z would only substantially change the
forecast if also using narrow bins for the Faint sample. We focus instead on
the Bright redshift uncertainty and fix the Faint photo-z to the fiducial
value.

Fig. \ref{errbiasphotoz} shows the FxB:All Bright bias error ratio between the
case of using 
spectroscopic (Bright) sample and the case of including redshift uncertainties
(photo-z) in the Bright sample. A vertical line (0.0035) show the expected photo-z
precision for the PAU narrow band survey \citeind{polpz}. The PAU photo-z would
for the Bright population recover 90\% of the bias error, with the Faint bias
being nearly unaffected. However, note that this conclusion depends strongly on
the Bright population redshift bin width, which fiducially is $\dzbin =
0.01(1+z)$ (see paper-I, section 3.3). Reducing the Bright redshift accuracy
from spec-z to a photo-z of $0.01(1+z)$ doubles the error on the Bright bias,
which illustrate how better photo-z than the typical bin width is important
when constraining cosmology with galaxy clustering.

The Bright bias ratios (Fig.\ref{errbiasphotoz}, top panel) approaches zero for
a typical photometric redshift. Analyzing the bright sample with 72 narrow
redshift bins and one bias parameter per bin is clearly not possible without
accurate redshifts. The Faint galaxy bias (bottom panel) is affected through the
Bright-Faint cross-correlations, sample variance cancellation and the
uncertainty in cosmological parameters. These effects are less direct and the
Faint bias error declines slower with increasing Bright photo-z, reaching
asymptotic values of $\sim$0.6-0.7 and larger for high redshifts. When
increasing the Bright photo-z, the Faint bias can still be measured through the
Faint clustering and counts-shear cross-correlations.

\xbf
\xfigure{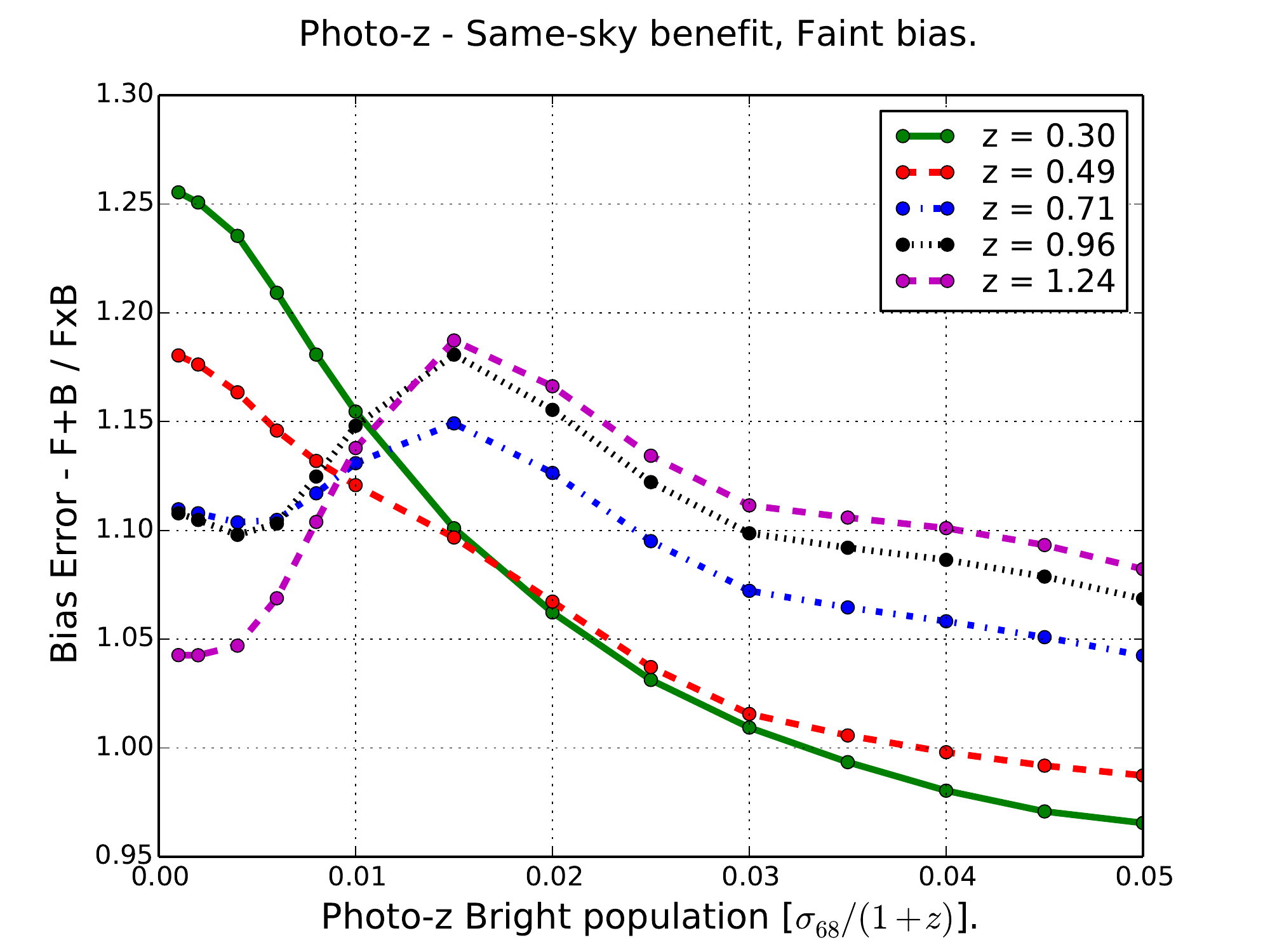}
%\xfigure{prod_biasgr21.pdf}
\caption{The same-sky bias error benefit when varying the Bright photo-z. Lines
are the Faint bias error ratios between non-overlapping and overlapping
surveys. On the x-axis is the Bright sample Gaussian photo-z (units of (1+z)).
The forecast include galaxy counts and shear (FxB:All).}
\label{errbias_photoz_samesky}
\xef

Fig. \ref{errbias_photoz_samesky} shows the same-sky ratio, which is the
bias error ratio between non-overlapping and overlapping galaxy surveys. Values
below unity means overlapping surveys improve the bias measurement. For both 
FxB and F+B, a higher photo-z result in higher bias errors and the absolute
errors increase strongly for all configurations (All, Counts). An increasing or decreasing
ratio results from errors growing faster for either overlapping or
non-overlapping surveys. For $\sigma_z \lesssim 0.01 (1+z)$ the Bright sample
alone becomes less able to constrain the bias, increasing the value of
overlapping samples. When the photo-z increase and the nearby bins becomes more
correlated, which reduce the same-sky benefit ratios. With only galaxy counts,
the same-sky ratio change less (not shown).

\subsection{Galaxy density}

\xbf
\xfigure{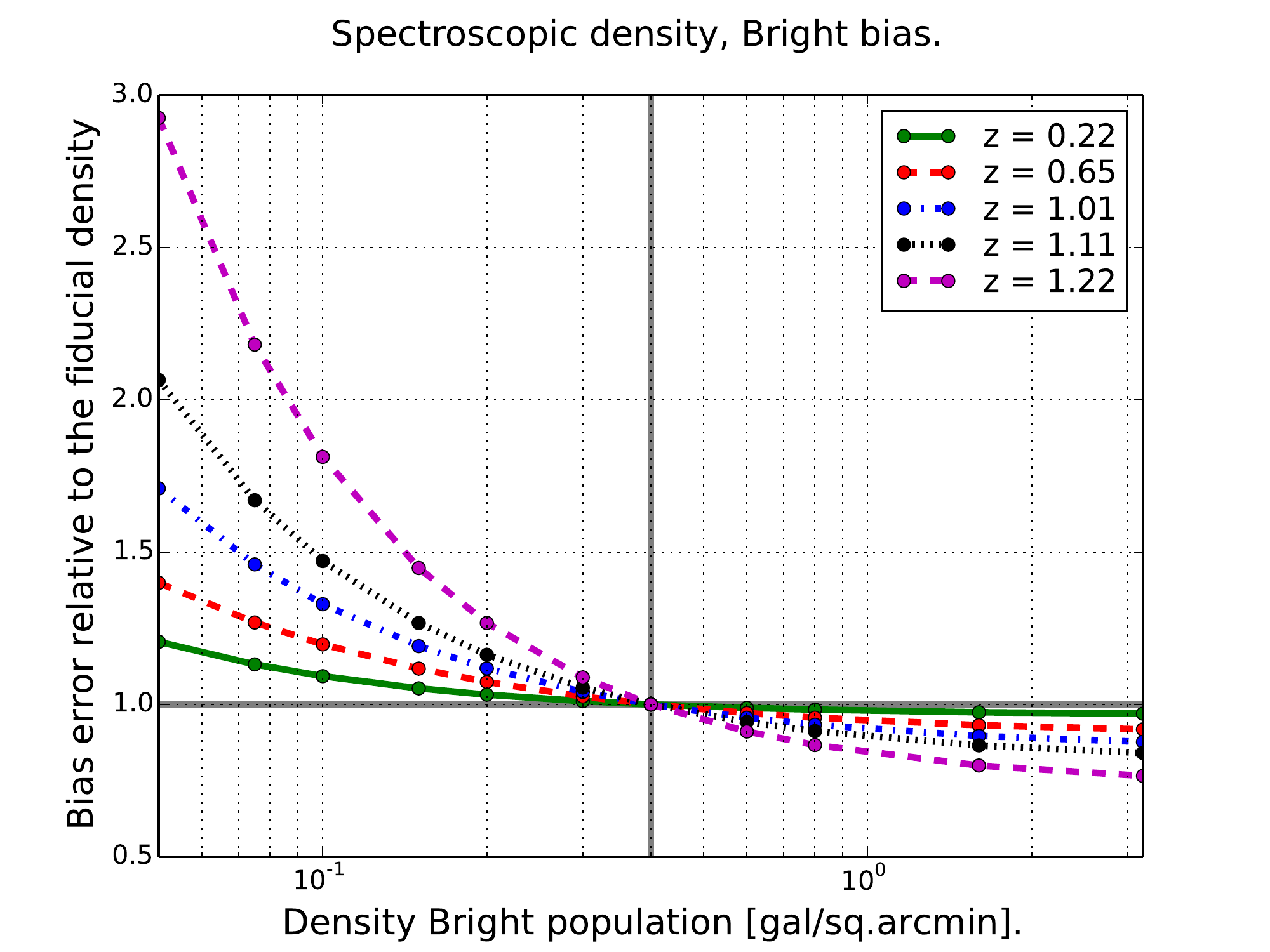}
\xfigure{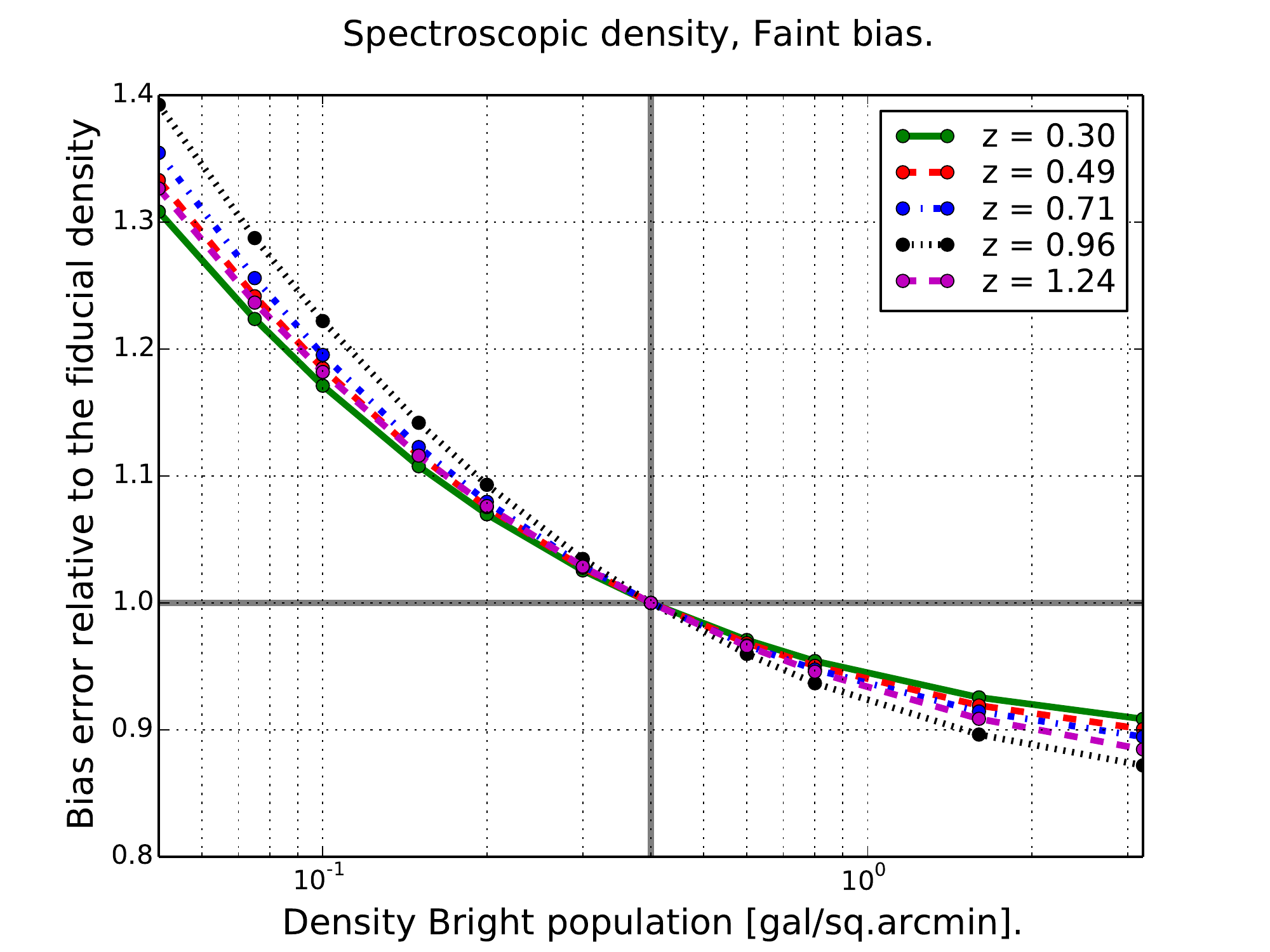}
\caption{Impact of the galaxy density on the bias error. Each line show the
FxB:All bias error, normalized to the errors for the fiducial Bright density
(0.4 gal/sq.arcmin) for different redshift bins (see legend). The top(bottom)
panel show the Bright(Faint) bias error ratio.}
\label{errbias_dens_ratio}
\xef

The galaxy density also change the cosmology and galaxy bias forecasts. This
subsection study the effect of the spectroscopic density directly on the bias
error. The fiducial photometric (Faint) sample is dense and varying its density
only introduce small changes in the bias error, although it impacts
cosmological constraint through e.g. the shear-shear measurement. For the
related FoM forecast, see \pssky.  

In Fig.\ref{errbias_dens_ratio} we show the bias error when varying the Bright
galaxy density and normalizing the error to unity for the fiducial density of the
Bright sample (0.4 gal/sq.arcmin). As expected, the bias error decrease
monotonically with higher galaxy density. The galaxy sample is less dense at
high redshifts, which leads to the error on bias improving most for the highest
redshift bins. When the Bright galaxy population is 0.1 gal/$\sqarcmin$, which
is $25\%$ of the fiducial value, the Bright bias error is 20-80$\%$ higher than
the fiducial error. The Faint bias (bottom panel) benefits indirect through 
Bright-Faint cross-correlations and sample variance cancellation, therefore
it depends weaker on the Bright galaxy density. Decreasing the Bright density 
to 0.1 gal/$\sqarcmin$ results in 20\% increase in the redshift error for most
redshift bins. When increasing the density beyond the fiducial value, both
populations show small improvements. This is compatible with the FoM saturation
around 0.4 gal/sq.arcmin in \pssky.

\bibliography{../exbib}{}

\begin{thebibliography}{56}
\expandafter\ifx\csname natexlab\endcsname\relax\def\natexlab#1{#1}\fi

\bibitem[{{Albrecht} {et~al}\mbox{.}(2006){Albrecht}, {Bernstein}, {Cahn},
  {Freedman}, {Hewitt}, {Hu}, {Huth}, {Kamionkowski}, {Kolb}, {Knox}, {Mather},
  {Staggs}, \& {Suntzeff}}]{detf}
{Albrecht} A. {et~al.}, 2006, ArXiv e-prints, 0609591

\bibitem[{{Anderson} {et~al}\mbox{.}(2014){Anderson}, {Aubourg}, {Bailey},
  {Beutler}, {Bolton}, {Brinkmann}, {Brownstein}, {Chuang}, {Cuesta}, {Dawson},
  {Eisenstein}, {Ho}, {Honscheid}, {Kazin}, {Kirkby}, {Manera}, {McBride},
  {Mena}, {Nichol}, {Olmstead}, {Padmanabhan}, {Palanque-Delabrouille},
  {Percival}, {Prada}, {Ross}, {Ross}, {S{\'a}nchez}, {Samushia}, {Schlegel},
  {Schneider}, {Seo}, {Strauss}, {Thomas}, {Tinker}, {Tojeiro}, {Verde},
  {Wake}, {Weinberg}, {Xu}, \& {Yeche}}]{andersonboss}
{Anderson} L. {et~al.}, 2014, \mnras, 439, 83

\bibitem[{{Anderson} {et~al}\mbox{.}(2012){Anderson}, {Aubourg}, {Bailey},
  {Bizyaev}, {Blanton}, {Bolton}, {Brinkmann}, {Brownstein}, {Burden},
  {Cuesta}, {da Costa}, {Dawson}, {de Putter}, {Eisenstein}, {Gunn}, {Guo},
  {Hamilton}, {Harding}, {Ho}, {Honscheid}, {Kazin}, {Kirkby}, {Kneib},
  {Labatie}, {Loomis}, {Lupton}, {Malanushenko}, {Malanushenko}, {Mandelbaum},
  {Manera}, {Maraston}, {McBride}, {Mehta}, {Mena}, {Montesano}, {Muna},
  {Nichol}, {Nuza}, {Olmstead}, {Oravetz}, {Padmanabhan},
  {Palanque-Delabrouille}, {Pan}, {Parejko}, {P{\^a}ris}, {Percival},
  {Petitjean}, {Prada}, {Reid}, {Roe}, {Ross}, {Ross}, {Samushia},
  {S{\'a}nchez}, {Schlegel}, {Schneider}, {Sc{\'o}ccola}, {Seo}, {Sheldon},
  {Simmons}, {Skibba}, {Strauss}, {Swanson}, {Thomas}, {Tinker}, {Tojeiro},
  {Maga{\~n}a}, {Verde}, {Wagner}, {Wake}, {Weaver}, {Weinberg}, {White}, {Xu},
  {Y{\`e}che}, {Zehavi}, \& {Zhao}}]{anderson_bao}
{Anderson} L. {et~al.}, 2012, \mnras, 427, 3435

\bibitem[{{Asorey}, {Crocce} \& {Gazta{\~n}aga}(2014){Asorey}, {Crocce}, \&
  {Gazta{\~n}aga}}]{asorey2}
{Asorey} J., {Crocce} M., {Gazta{\~n}aga} E., 2014, \mnras, 445, 2825

\bibitem[{{Asorey} {et~al}\mbox{.}(2012){Asorey}, {Crocce}, {Gazta{\~n}aga}, \&
  {Lewis}}]{asorey}
{Asorey} J., {Crocce} M., {Gazta{\~n}aga} E., {Lewis} A., 2012, \mnras, 427,
  1891

\bibitem[{{Bardeen} {et~al}\mbox{.}(1986){Bardeen}, {Bond}, {Kaiser}, \&
  {Szalay}}]{clustering_bardeen}
{Bardeen} J.~M., {Bond} J.~R., {Kaiser} N., {Szalay} A.~S., 1986, \apj, 304, 15

\bibitem[{{Bel}, {Hoffmann} \& {Gazta{\~n}aga}(2015){Bel}, {Hoffmann}, \&
  {Gazta{\~n}aga}}]{bel2015}
{Bel} J., {Hoffmann} K., {Gazta{\~n}aga} E., 2015, ArXiv e-prints

\bibitem[{{Bernstein} \& {Cai}(2011)}]{cai2011}
{Bernstein} G.~M., {Cai} Y.-C., 2011, \mnras, 416, 3009

\bibitem[{{Blanton} {et~al}\mbox{.}(2000){Blanton}, {Cen}, {Ostriker},
  {Strauss}, \& {Tegmark}}]{galaxyev_blanton}
{Blanton} M., {Cen} R., {Ostriker} J.~P., {Strauss} M.~A., {Tegmark} M., 2000,
  \apj, 531, 1

\bibitem[{{Cacciato} {et~al}\mbox{.}(2013){Cacciato}, {van den Bosch}, {More},
  {Mo}, \& {Yang}}]{cacciato_hod}
{Cacciato} M., {van den Bosch} F.~C., {More} S., {Mo} H., {Yang} X., 2013,
  \mnras, 430, 767

\bibitem[{{Cai} \& {Bernstein}(2012)}]{cai2012}
{Cai} Y.-C., {Bernstein} G., 2012, \mnras, 422, 1045

\bibitem[{{Clerkin} {et~al}\mbox{.}(2015){Clerkin}, {Kirk}, {Lahav}, {Abdalla},
  \& {Gazta{\~n}aga}}]{clerkin}
{Clerkin} L., {Kirk} D., {Lahav} O., {Abdalla} F.~B., {Gazta{\~n}aga} E., 2015,
  \mnras, 448, 1389

\bibitem[{{Contreras} {et~al}\mbox{.}(2013){Contreras}, {Blake}, {Poole},
  {Marin}, {Brough}, {Colless}, {Couch}, {Croom}, {Croton}, {Davis},
  {Drinkwater}, {Forster}, {Gilbank}, {Gladders}, {Glazebrook}, {Jelliffe},
  {Jurek}, {Li}, {Madore}, {Martin}, {Pimbblet}, {Pracy}, {Sharp}, {Wisnioski},
  {Woods}, {Wyder}, \& {Yee}}]{wigglez}
{Contreras} C. {et~al.}, 2013, \mnras, 430, 924

\bibitem[{{Coupon} {et~al}\mbox{.}(2012){Coupon}, {Kilbinger}, {McCracken},
  {Ilbert}, {Arnouts}, {Mellier}, {Abbas}, {de la Torre}, {Goranova},
  {Hudelot}, {Kneib}, \& {Le F{\`e}vre}}]{couponhod}
{Coupon} J. {et~al.}, 2012, \aap, 542, A5

\bibitem[{{de Jong} {et~al}\mbox{.}(2013){de Jong}, {Verdoes Kleijn},
  {Kuijken}, \& {Valentijn}}]{kids}
{de Jong} J.~T.~A., {Verdoes Kleijn} G.~A., {Kuijken} K.~H., {Valentijn} E.~A.,
  2013, Experimental Astronomy, 35, 25

\bibitem[{{de la Torre} {et~al}\mbox{.}(2013){de la Torre}, {Guzzo}, {Peacock},
  {Branchini}, {Iovino}, {Granett}, {Abbas}, {Adami}, {Arnouts}, {Bel},
  {Bolzonella}, {Bottini}, {Cappi}, {Coupon}, {Cucciati}, {Davidzon}, {De
  Lucia}, {Fritz}, {Franzetti}, {Fumana}, {Garilli}, {Ilbert}, {Krywult}, {Le
  Brun}, {Le F{\`e}vre}, {Maccagni}, {Ma{\l}ek}, {Marulli}, {McCracken},
  {Moscardini}, {Paioro}, {Percival}, {Polletta}, {Pollo}, {Schlagenhaufer},
  {Scodeggio}, {Tasca}, {Tojeiro}, {Vergani}, {Zanichelli}, {Burden}, {Di
  Porto}, {Marchetti}, {Marinoni}, {Mellier}, {Monaco}, {Nichol}, {Phleps},
  {Wolk}, \& {Zamorani}}]{vipers}
{de la Torre} S. {et~al.}, 2013, \aap, 557, A54

\bibitem[{{de Putter}, {Dor{\'e}} \& {Takada}(2013){de Putter}, {Dor{\'e}}, \&
  {Takada}}]{deputter}
{de Putter} R., {Dor{\'e}} O., {Takada} M., 2013, preprint (arXiv:1308.6070)

\bibitem[{{Dekel} \& {Lahav}(1999)}]{stoch_bias_var}
{Dekel} A., {Lahav} O., 1999, \apj, 520, 24

\bibitem[{{Di Dio} {et~al}\mbox{.}(2013){Di Dio}, {Montanari}, {Durrer}, \&
  {Lesgourgues}}]{dionbins}
{Di Dio} E., {Montanari} F., {Durrer} R., {Lesgourgues} J., 2013, ArXiv
  e-prints

\bibitem[{{Eisenstein} {et~al}\mbox{.}(2007){Eisenstein}, {Seo}, {Sirko}, \&
  {Spergel}}]{eisenstein_bao1}
{Eisenstein} D.~J., {Seo} H.-J., {Sirko} E., {Spergel} D.~N., 2007, \apj, 664,
  675

\bibitem[{{Eisenstein}, {Seo} \& {White}(2007){Eisenstein}, {Seo}, \&
  {White}}]{eisenstein_bao2}
{Eisenstein} D.~J., {Seo} H.-J., {White} M., 2007, \apj, 664, 660

\bibitem[{{Eisenstein} {et~al}\mbox{.}(2005){Eisenstein}, {Zehavi}, {Hogg},
  {Scoccimarro}, {Blanton}, {Nichol}, {Scranton}, {Seo}, {Tegmark}, {Zheng},
  {Anderson}, {Annis}, {Bahcall}, {Brinkmann}, {Burles}, {Castander},
  {Connolly}, {Csabai}, {Doi}, {Fukugita}, {Frieman}, {Glazebrook}, {Gunn},
  {Hendry}, {Hennessy}, {Ivezi{\'c}}, {Kent}, {Knapp}, {Lin}, {Loh}, {Lupton},
  {Margon}, {McKay}, {Meiksin}, {Munn}, {Pope}, {Richmond}, {Schlegel},
  {Schneider}, {Shimasaku}, {Stoughton}, {Strauss}, {SubbaRao}, {Szalay},
  {Szapudi}, {Tucker}, {Yanny}, \& {York}}]{eisenstein2005}
{Eisenstein} D.~J. {et~al.}, 2005, \apj, 633, 560

\bibitem[{{Eriksen} \& {Gaztanaga}(2014{\natexlab{a}})}]{sameskyX}
{Eriksen} M., {Gaztanaga} E., 2014{\natexlab{a}}, (arXiv:1412.8429) (EG14a)

\bibitem[{{Eriksen} \& {Gaztanaga}(2014{\natexlab{b}})}]{paperI}
{Eriksen} M., {Gaztanaga} E., 2014{\natexlab{b}}, (arXiv:1412.2208) (paper-I)

\bibitem[{{Eriksen} \& {Gaztanaga}(2015)}]{paperII}
{Eriksen} M., {Gaztanaga} E., 2015, (arXiv:1502.03972) (paper-II)

\bibitem[{{Font-Ribera} {et~al}\mbox{.}(2014){Font-Ribera}, {McDonald},
  {Mostek}, {Reid}, {Seo}, \& {Slosar}}]{mcdonald}
{Font-Ribera} A., {McDonald} P., {Mostek} N., {Reid} B.~A., {Seo} H.-J.,
  {Slosar} A., 2014, \jcap, 5, 23

\bibitem[{{Frieman} \& {Gaztanaga}(1994)}]{yet3p}
{Frieman} J.~A., {Gaztanaga} E., 1994, \apj, 425, 392

\bibitem[{{Fry}(1996)}]{galaxyev2}
{Fry} J.~N., 1996, \apjl, 461, L65

\bibitem[{{Gazta{\~n}aga}, {Cabr{\'e}} \& {Hui}(2009){Gazta{\~n}aga},
  {Cabr{\'e}}, \& {Hui}}]{cabreplot}
{Gazta{\~n}aga} E., {Cabr{\'e}} A., {Hui} L., 2009, \mnras, 399, 1663

\bibitem[{{Gazta{\~n}aga} {et~al}\mbox{.}(2012){Gazta{\~n}aga}, {Eriksen},
  {Crocce}, {Castander}, {Fosalba}, {Marti}, {Miquel}, \& {Cabr{\'e}}}]{gazta}
{Gazta{\~n}aga} E., {Eriksen} M., {Crocce} M., {Castander} F.~J., {Fosalba} P.,
  {Marti} P., {Miquel} R., {Cabr{\'e}} A., 2012, \mnras, 422, 2904

\bibitem[{{Gazta{\~n}aga} {et~al}\mbox{.}(2005){Gazta{\~n}aga}, {Norberg},
  {Baugh}, \& {Croton}}]{3ptmes}
{Gazta{\~n}aga} E., {Norberg} P., {Baugh} C.~M., {Croton} D.~J., 2005, \mnras,
  364, 620

\bibitem[{{Gazta{\~n}aga} \& {Scoccimarro}(2005)}]{3ptfunction}
{Gazta{\~n}aga} E., {Scoccimarro} R., 2005, \mnras, 361, 824

\bibitem[{{Heymans} {et~al}\mbox{.}(2013){Heymans}, {Grocutt}, {Heavens},
  {Kilbinger}, {Kitching}, {Simpson}, {Benjamin}, {Erben}, {Hildebrandt},
  {Hoekstra}, {Mellier}, {Miller}, {Van Waerbeke}, {Brown}, {Coupon}, {Fu},
  {Harnois-D{\'e}raps}, {Hudson}, {Kuijken}, {Rowe}, {Schrabback}, {Semboloni},
  {Vafaei}, \& {Velander}}]{heymans_lens}
{Heymans} C. {et~al.}, 2013, \mnras, 432, 2433

\bibitem[{{Ivezic} {et~al}\mbox{.}(2008){Ivezic}, {Tyson}, {Acosta}, {Allsman},
  {Anderson}, {Andrew}, {Angel}, {Axelrod}, {Barr}, {Becker}, {Becla},
  {Beldica}, {Blandford}, {Bloom}, {Borne}, {Brandt}, {Brown}, {Bullock},
  {Burke}, {Chandrasekharan}, {Chesley}, {Claver}, {Connolly}, {Cook},
  {Cooray}, {Covey}, {Cribbs}, {Cutri}, {Daues}, {Delgado}, {Ferguson},
  {Gawiser}, {Geary}, {Gee}, {Geha}, {Gibson}, {Gilmore}, {Gressler}, {Hogan},
  {Huffer}, {Jacoby}, {Jain}, {Jernigan}, {Jones}, {Juric}, {Kahn}, {Kalirai},
  {Kantor}, {Kessler}, {Kirkby}, {Knox}, {Krabbendam}, {Krughoff}, {Kulkarni},
  {Lambert}, {Levine}, {Liang}, {Lim}, {Lupton}, {Marshall}, {Marshall}, {May},
  {Miller}, {Mills}, {Monet}, {Neill}, {Nordby}, {O'Connor}, {Oliver},
  {Olivier}, {Olsen}, {Owen}, {Peterson}, {Petry}, {Pierfederici},
  {Pietrowicz}, {Pike}, {Pinto}, {Plante}, {Radeka}, {Rasmussen}, {Ridgway},
  {Rosing}, {Saha}, {Schalk}, {Schindler}, {Schneider}, {Schumacher}, {Sebag},
  {Seppala}, {Shipsey}, {Silvestri}, {Smith}, {Smith}, {Strauss}, {Stubbs},
  {Sweeney}, {Szalay}, {Thaler}, {Vanden Berk}, {Walkowicz}, {Warner},
  {Willman}, {Wittman}, {Wolff}, {Wood-Vasey}, {Yoachim}, {Zhan}, \& {for the
  LSST Collaboration}}]{lsst1}
{Ivezic} Z. {et~al.}, 2008, ArXiv e-prints

\bibitem[{{Kaiser}(1984)}]{clustering_kaiser}
{Kaiser} N., 1984, \apjl, 284, L9

\bibitem[{{Kirk} {et~al}\mbox{.}(2013){Kirk}, {Lahav}, {Bridle}, {Jouvel},
  {Abdalla}, \& {Frieman}}]{kirk}
{Kirk} D., {Lahav} O., {Bridle} S., {Jouvel} S., {Abdalla} F.~B., {Frieman}
  J.~A., 2013, preprint (arXiv:1307.8062)

\bibitem[{{Laureijs} {et~al}\mbox{.}(2011){Laureijs}, {Amiaux}, {Arduini},
  {Augu{\`e}res}, {Brinchmann}, {Cole}, {Cropper}, {Dabin}, {Duvet}, {Ealet},
  \& et~al.}]{euclid}
{Laureijs} R. {et~al.}, 2011, preprint (arXiv:1110.3193)

\bibitem[{{Le F{\`e}vre} {et~al}\mbox{.}(2013){Le F{\`e}vre}, {Cassata},
  {Cucciati}, {Garilli}, {Ilbert}, {Le Brun}, {Maccagni}, {Moreau},
  {Scodeggio}, {Tresse}, {Zamorani}, {Adami}, {Arnouts}, {Bardelli},
  {Bolzonella}, {Bondi}, {Bongiorno}, {Bottini}, {Cappi}, {Charlot}, {Ciliegi},
  {Contini}, {de la Torre}, {Foucaud}, {Franzetti}, {Gavignaud}, {Guzzo},
  {Iovino}, {Lemaux}, {L{\'o}pez-Sanjuan}, {McCracken}, {Marano}, {Marinoni},
  {Mazure}, {Mellier}, {Merighi}, {Merluzzi}, {Paltani}, {Pell{\`o}}, {Pollo},
  {Pozzetti}, {Scaramella}, {Tasca}, {Vergani}, {Vettolani}, {Zanichelli}, \&
  {Zucca}}]{vvds}
{Le F{\`e}vre} O. {et~al.}, 2013, \aap, 559, A14

\bibitem[{{Levi} {et~al}\mbox{.}(2013){Levi}, {Bebek}, {Beers}, {Blum}, {Cahn},
  {Eisenstein}, {Flaugher}, {Honscheid}, {Kron}, {Lahav}, {McDonald}, {Roe},
  {Schlegel}, \& {representing the DESI collaboration}}]{msdesi}
{Levi} M. {et~al.}, 2013, preprint (ArXiv:1308.0847)

\bibitem[{{Mart{\'{\i}}} {et~al}\mbox{.}(2014){Mart{\'{\i}}}, {Miquel},
  {Castander}, {Gazta{\~n}aga}, {Eriksen}, \& {S{\'a}nchez}}]{polpz}
{Mart{\'{\i}}} P., {Miquel} R., {Castander} F.~J., {Gazta{\~n}aga} E.,
  {Eriksen} M., {S{\'a}nchez} C., 2014, MNRAS, 442, 92

\bibitem[{{McDonald} \& {Seljak}(2009)}]{mcdonseljak}
{McDonald} P., {Seljak} U., 2009, \jcap, 10, 7

\bibitem[{{Nock}, {Percival} \& {Ross}(2010){Nock}, {Percival}, \&
  {Ross}}]{rsdin2d}
{Nock} K., {Percival} W.~J., {Ross} A.~J., 2010, \mnras, 407, 520

\bibitem[{{Nusser} \& {Davis}(1994)}]{galaxyev1}
{Nusser} A., {Davis} M., 1994, \apjl, 421, L1

\bibitem[{{Perlmutter} {et~al}\mbox{.}(1999){Perlmutter}, {Aldering},
  {Goldhaber}, {Knop}, {Nugent}, {Castro}, {Deustua}, {Fabbro}, {Goobar},
  {Groom}, {Hook}, {Kim}, {Kim}, {Lee}, {Nunes}, {Pain}, {Pennypacker},
  {Quimby}, {Lidman}, {Ellis}, {Irwin}, {McMahon}, {Ruiz-Lapuente}, {Walton},
  {Schaefer}, {Boyle}, {Filippenko}, {Matheson}, {Fruchter}, {Panagia},
  {Newberg}, {Couch}, \& {Project}}]{perlmutterSN}
{Perlmutter} S. {et~al.}, 1999, \apj, 517, 565

\bibitem[{{Planck Collaboration} {et~al}\mbox{.}(2015){Planck Collaboration},
  {Ade}, {Aghanim}, {Arnaud}, {Ashdown}, {Aumont}, {Baccigalupi}, {Banday},
  {Barreiro}, {Bartlett}, \& et~al.}]{planck2015}
{Planck Collaboration} {et~al.}, 2015, preprint (ArXiv:1502.01589)

\bibitem[{{Press} \& {Schechter}(1974)}]{galform}
{Press} W.~H., {Schechter} P., 1974, \apj, 187, 425

\bibitem[{{Riess} {et~al}\mbox{.}(1998){Riess}, {Filippenko}, {Challis},
  {Clocchiatti}, {Diercks}, {Garnavich}, {Gilliland}, {Hogan}, {Jha},
  {Kirshner}, {Leibundgut}, {Phillips}, {Reiss}, {Schmidt}, {Schommer},
  {Smith}, {Spyromilio}, {Stubbs}, {Suntzeff}, \& {Tonry}}]{riessSN}
{Riess} A.~G. {et~al.}, 1998, \aj, 116, 1009

\bibitem[{{Scoccimarro} {et~al}\mbox{.}(2001){Scoccimarro}, {Sheth}, {Hui}, \&
  {Jain}}]{hod1}
{Scoccimarro} R., {Sheth} R.~K., {Hui} L., {Jain} B., 2001, \apj, 546, 20

\bibitem[{{Sefusatti} {et~al}\mbox{.}(2006){Sefusatti}, {Crocce}, {Pueblas}, \&
  {Scoccimarro}}]{another3pt}
{Sefusatti} E., {Crocce} M., {Pueblas} S., {Scoccimarro} R., 2006, \prd, 74,
  023522

\bibitem[{{Seljak} \& {Warren}(2004)}]{biasstoch}
{Seljak} U., {Warren} M.~S., 2004, \mnras, 355, 129

\bibitem[{{Shoji}, {Jeong} \& {Komatsu}(2009){Shoji}, {Jeong}, \&
  {Komatsu}}]{fullcls}
{Shoji} M., {Jeong} D., {Komatsu} E., 2009, \apj, 693, 1404

\bibitem[{{Simpson} {et~al}\mbox{.}(2013){Simpson}, {Heymans}, {Parkinson},
  {Blake}, {Kilbinger}, {Benjamin}, {Erben}, {Hildebrandt}, {Hoekstra},
  {Kitching}, {Mellier}, {Miller}, {Van Waerbeke}, {Coupon}, {Fu},
  {Harnois-D{\'e}raps}, {Hudson}, {Kuijken}, {Rowe}, {Schrabback}, {Semboloni},
  {Vafaei}, \& {Velander}}]{simpsonWLRSD}
{Simpson} F. {et~al.}, 2013, \mnras, 429, 2249

\bibitem[{{Tegmark} \& {Peebles}(1998)}]{galaxyev3}
{Tegmark} M., {Peebles} P.~J.~E., 1998, \apjl, 500, L79

\bibitem[{{The Dark Energy Survey Collaboration}(2005)}]{des}
{The Dark Energy Survey Collaboration}, 2005, preprint (arXiv:astro-ph/0510346)

\bibitem[{{Weinberg} {et~al}\mbox{.}(2013){Weinberg}, {Mortonson},
  {Eisenstein}, {Hirata}, {Riess}, \& {Rozo}}]{weinberg}
{Weinberg} D.~H., {Mortonson} M.~J., {Eisenstein} D.~J., {Hirata} C., {Riess}
  A.~G., {Rozo} E., 2013, \physrep, 530, 87

\bibitem[{{Zheng} {et~al}\mbox{.}(2005){Zheng}, {Berlind}, {Weinberg},
  {Benson}, {Baugh}, {Cole}, {Dav{\'e}}, {Frenk}, {Katz}, \&
  {Lacey}}]{zhenghod}
{Zheng} Z. {et~al.}, 2005, \apj, 633, 791

\end{thebibliography}
\bibliographystyle{mn2e}
\end{document}